\documentclass[12pt,preprint]{aastex}
\def\kms{{\rm km\,s^{-1}}}
\def\kmsd{{\rm km\,s^{-1}\,deg^{-1}}}
\def\kpc{{\rm kpc}}

\def\masyr{{\rm mas}\,{\rm yr}^{-1}}
\def\muas{{\mu\rm as}}
\def\lim{{\rm lim}}

\def\bv{{\bf v}}
\newcommand{\bdv}[1]{\mbox{\boldmath$#1$}}
\def\e{{\rm E}}
\def\bpi{{\bdv{\pi}}}
\def\balpha{{\bdv{\alpha}}}
\def\bLambda{{\bdv{\Lambda}}}
\def\bvt{{\bdv{\tilde v}}}

\def\rel{{\rm rel}}

\def\au{{\rm AU}}

\def\bmu{{\bdv{\mu}}}

\def\kms{{\rm km}\,{\rm s}^{-1}}
\def\l{{\rm L}}
\def\s{{\rm S}}
\def\ls{{\rm LS}}
\def\alphasub{{\phi}}

\newcommand{\grad}{\hbox{$^\circ$}}
\begin{document}

\title{First Space-Based Microlens Parallax Measurement:
{\it Spitzer} Observations of OGLE-2005-SMC-001}

\author{
Subo Dong\altaffilmark{1,2},
A.~Udalski\altaffilmark{3,4},
A.~Gould\altaffilmark{1,2},
W.T.~Reach\altaffilmark{5},
G.W.~Christie\altaffilmark{2,6},
A.F.~Boden\altaffilmark{7,8},
D.P.~Bennett\altaffilmark{9},
G.~Fazio\altaffilmark{10}
K.~Griest\altaffilmark{11}
M.K.~Szyma{\'n}ski\altaffilmark{3,4},
M.~Kubiak\altaffilmark{3,4},
I.~Soszy{\'n}ski\altaffilmark{3,4},
G.~Pietrzy{\'n}ski\altaffilmark{3,4,12},
O.~Szewczyk\altaffilmark{3,4},
{\L}.~Wyrzykowski\altaffilmark{3,4,13},
{K}.~Ulaczyk\altaffilmark{3,4},
T.~Wieckowski\altaffilmark{3,4},
B.~Paczy{\'n}ski\altaffilmark{3,14}
D.L.~DePoy\altaffilmark{1,2},
R.W.~Pogge\altaffilmark{1,2}
G.W.~Preston\altaffilmark{15}
I.B.~Thompson\altaffilmark{15}
B.M.~Patten\altaffilmark{10}
}
\altaffiltext{1}
{Department of Astronomy, Ohio State University,
140 W.\ 18th Ave., Columbus, OH 43210, USA; 
depoy,dong,gould,pogge@astronomy.ohio-state.edu}
\altaffiltext{2}
{Microlensing Follow Up Network ($\mu$FUN)}
\altaffiltext{3}
{Optical Gravitational Lens Experiment (OGLE)}
\altaffiltext{4}
{Warsaw University Observatory, Al.~Ujazdowskie~4, 00-478~Warszawa, Poland; 
udalski, msz, mk, soszynsk, pietrzyn, szewczyk, wyrzykow, kulaczyk, twieck@astrouw.edu.pl}
\altaffiltext{5}
{{\it Spitzer} Science Center, California Institute of Technology,
Pasadena, CA 91125, USA; reach@ipac.caltech.edu}
\altaffiltext{6}
{Auckland Observatory, Auckland, New Zealand, gwchristie@christie.org.nz}
\altaffiltext{7}
{Michelson Science Center, California Institute of Technology,
770 South Wilson Ave., Pasadena, CA 91125, USA; bode@ipac.caltech.edu}
\altaffiltext{8}
{Department of Physics and Astronomy, Georgia State University,
29 Peachtree Center Avenue, Suite 400, Atlanta, GA 30303, USA.}
\altaffiltext{9}
{Department of Physics, Notre Dame University, Notre Dame, IN 46556, USA;
bennett@nd.edu}
\altaffiltext{10}
{Center for Astrophysics, Cambridge, MA 02138, USA; fazio,bpatten@cfa.harvard.edu}
\altaffiltext{11}
{Department of Physics, University of California, San Diego, CA 92093, USA;
griest@ucsd.edu}
\altaffiltext{12}{Universidad de Concepci{\'o}n, Departamento de Fisica,
Casilla 160--C, Concepci{\'o}n, Chile}
\altaffiltext{13} {Institute of Astronomy  Cambridge University,
Madingley Rd., CB3 0HA Cambridge}
\altaffiltext{14}{Princeton University Observatory, Princeton, NJ 08544, 
USA; bp@astro.princeton.edu}
\altaffiltext{15}
{The Observatories of the Carnegie Institute of Washington, 813 Santa Barbara
Street, Pasadena, CA 91101, USA; gwp,ian@ociw.edu}

\begin{abstract}
We combine {\it Spitzer} and ground-based observations to measure
the microlens parallax of OGLE-2005-SMC-001, the first 
such space-based determination since S.~Refsdal proposed the idea in 1966.
The parallax measurement yields a projected velocity 
$\tilde v\sim 230\,\kms$, the typical value expected for halo lenses,
but an order of magnitude smaller than would be expected for lenses
lying in the Small Magellanic Cloud (SMC) itself.  The lens is a weak
(i.e., non-caustic-crossing) binary, which complicates the analysis
considerably but ultimately contributes additional constraints.
Using a test proposed by Assef et al. (2006), which makes use
only of kinematic information about different populations but
does not make any assumptions about their respective mass functions,
we find that the likelihood ratio is 
${\cal L}_{\rm halo}/{\cal L}_{\rm SMC}=20$.  Hence, halo lenses are
strongly favored but SMC lenses are not definitively ruled out.  Similar
{\it Spitzer} observations of additional lenses toward the Magellanic Clouds
would clarify the nature of the lens population.  The {\it Space
Interferometry Mission} could make even more constraining measurements.

\end{abstract}
 
\section{Introduction
\label{sec:intro}}

In a visionary paper written more than 40 years ago, \citet{refsdal66} argued
that two important but otherwise unmeasurable parameters of microlensing 
events could be determined by simultaneously observing the event from 
the Earth and a satellite in solar orbit.  In modern language, these
are the Einstein radius projected onto the observer plane, $\tilde r_\e$,
and the direction of lens-source relative proper motion.  Since the
Einstein timescale, $t_\e$, is routinely measured for all events, these
parameter determinations are equivalent to knowing the projected 
relative
velocity,
$\bvt$, whose magnitude is simply $\tilde v\equiv \tilde r_\e/t_\e$.
Here, $\tilde r_\e\equiv\au/\pi_\e$, $t_\e=\theta_\e/\mu$, and
\begin{equation}
\pi_\e = \sqrt{\pi_\rel\over\kappa M},\quad
\theta_\e = \sqrt{\kappa M\pi_\rel},
\label{eqn:piedef}
\end{equation}
where $\pi_\e$ is the microlens parallax, $M$ is the mass of the lens,
$\theta_\e$ is the angular Einstein radius,
$\pi_\rel$ and $\bmu$ are the lens-source relative parallax and
proper motion, respectively, and $\kappa\equiv 4 G/(c^2 \au)$.

The practical importance of this suggestion became clear when the
MACHO \citep{alcock93} and EROS \citep{aubourg93} collaborations
reported the detection
of microlensing events toward the Large Magellanic Cloud (LMC).
Over the course of time, MACHO \citep{alcock97,alcock00} has found about
15 such events and argued that these imply that about 20\% of the
Milky Way dark halo is composed of compact objects (``MACHOs''),
while EROS \citep{afonso03,tisserand06} has argued that their relative
lack of such detections was consistent with all the events being
due to stars in the Milky Way disk or the Magellanic Clouds (MCs) themselves.
For any given individual event, it is generally impossible to tell
(with only a measurement of $t_\e$)
where along the line of sight the lens lies, so one cannot distinguish
among the three possibilities: Milky Way disk, Milky Way halo, or
``self-lensing'' in which the source and lens both lie in the same
external galaxy.

However, as \citet{boutreux96} argued, measurement of $\bvt$ might allow 
one to distinguish among these populations with good confidence:
disk, halo, and MC lenses typically have $\tilde v$ values of
$50$, $300$, and $2000\,\kms$, respectively.  The
high projected speed of MC lenses derives from the long ``lever
arm'' that multiplies their small local transverse speed by
the ratio of the distances from the observer and the lens to the source.

There are serious obstacles, both practical and theoretical to
measuring $\bvt$.  One obvious practical problem is simply
launching a spacecraft with a suitable camera into solar orbit.
But the theoretical difficulties also place significant constraints
on the characteristics of that spacecraft.  To understand these
properly, one should think in terms of the ``microlens parallax''
$\bpi_\e$, whose magnitude is $\pi_\e\equiv \au/\tilde r_\e$ and whose
direction is the same as $\bvt$.  Choosing a coordinate system whose
$x$-axis is aligned with the Earth-satellite separation at the peak of the
event, we can write $\bpi_\e = (\pi_{\e,\tau},\pi_{\e,\beta})$.  Then
to good approximation,
\begin{equation}
\bpi_\e = (\pi_{\e,\tau},\pi_{\e,\beta}) = 
{\au\over d_\perp}\biggl({\Delta t_0\over t_\e},\Delta u_0\biggr),
\label{eqn:pitaubeta}
\end{equation}
where $d_\perp$ is the Earth-satellite separation (projected onto
the plane of the sky),
$\Delta t_0$ is the difference in time of event maximum as seen
from the Earth and satellite, and $\Delta u_0$ is the difference in
dimensionless impact parameter (determined from the maximum observed
magnification).  

\citet{refsdal66} already realized that equation (\ref{eqn:pitaubeta})
implicitly contains a four-fold degeneracy: while $\Delta t_0/t_\e$
is unambiguously determined, there are four different values 
of $\Delta u_0$ that depend on whether the individual impact parameters
are positive or negative (on one side of the lens or the other;
see Fig. 2 of \citealt{gould94}).  In fact, the situation is considerably
worse than this.  While $t_0$ is usually measured very precisely
in individual microlensing events, $u_0$ typically has much larger
errors because it is strongly correlated with three other parameters,
the timescale, $t_\e$, the source flux $f_{\rm s}$, and the blended flux,
$f_{\rm b}$.
For a satellite separated by $d_\perp \sim 0.2\,\au$, and a projected
Einstein radius $\tilde r_\e \sim 5\,\au$, errors in the impact-parameter
determinations of only $\sigma(u_0)\sim 2\%$ would lead to fractional
errors $\sigma(\pi_\e)/\pi_\e \sim \sqrt{2}\sigma(u_0)\tilde r_\e/d_\perp
\sim 70\%$.  However, \citet{gould95} showed that if the two cameras
had essentially identical spectral responses and similar 
point-spread-functions, so that one knew a priori that the blended light
was virtually identical for the Earth and satellite measurements, then
the error in $\Delta u_0$ would be reduced far below the individual
errors in $u_0$, making the parallax determination once again feasible.

Unfortunately, this trick cannot be used on {\it Spitzer}, the first 
general purpose camera to be placed in solar orbit.  The shortest
wavelength at which {\it Spitzer} operates is the $L$ band ($3.6\,\mu$m),
implying that the camera's sensitivity cannot be duplicated from the
ground,
because of both higher background and different throughput as a function
of wavelength.

In principle, microlens parallaxes can also be measured from the ground.
As with space-based parallaxes, one component of $\bpi_\e$ can generally
be measured much more precisely than the other.  For most events,
$t_\e\ll {\rm yr}$, and for these the Earth's acceleration can
be approximated as constant during the event.  To the degree that 
this acceleration is aligned (anti-aligned) with the 
lens-source relative motion, 
it induces an asymmetry in the light curve, since the event proceeds
faster (slower) before peak than afterward \citep{gmb94}.  This
is characterized by the ``asymmetry parameter'' 
$\Gamma\equiv\bpi_\e\cdot\balpha=\pi_{\e,\parallel}\alpha$, where
$\balpha$ is the apparent acceleration of the Sun projected onto
the sky and normalized to an AU, and $\pi_{\e,\parallel}$ is the
component of $\bpi_\e$ parallel to $\balpha$.  Since $\Gamma$ is
directly measurable from the light curve, one can directly obtain
1-D parallax information $\pi_{\e,\parallel}=\Gamma/\alpha$
for these events, while the orthogonal component $\pi_{\e,\perp}$
is measured extremely poorly (e.g., \citealt{ghosh05,jiang05}).
While there are a few exceptions \citep{alcock01,gba04,park04},
2-D parallaxes can generally only be obtained for relatively
long events $t_\e\ga 90\,\rm days$, and even for these, the
$\bpi_\e$ error ellipse is generally elongated in the $\pi_{\e,\perp}$ 
direction \citep{poindexter05}.

Since {\it Spitzer} is in an Earth-trailing orbit
and the SMC is close to the ecliptic pole, 
the $\pi_{E,\tau}$
direction (defined by the Earth-satellite separation vector) is
very nearly orthogonal to the $\pi_{\e,\parallel}$ direction
(defined by the direction of the Sun).  Recognizing this, \citet{gould99}
advocated combining the two essentially 1-D parallaxes from the 
Earth-{\it Spitzer} comparison and the accelerating Earth alone to
produce a single 2-D measurement of $\bpi_\e$.  He noted that once
the difficult problem of measuring $\pi_{\e,\beta}$ was jettisoned,
the satellite observations could be streamlined to a remarkable degree:
essentially only 3 observations were needed, 2 at times placed
symmetrically around the peak, which are sensitive to the offset in
$t_0$ between the Earth and satellite, and a third at late times
to set the flux scale.  This streamlining is important
from a practical point of view because Target-of-Opportunity (ToO) time
on {\it Spitzer} incurs a large penalty.  \citet{gould99} noted that
the components of $\bpi_\e$ measured by the two techniques were not
exactly orthogonal but argued (incorrectly as it turns out) that
this had no significant consequences for the experiment.  We return
to this point below.

Here we analyze {\it Spitzer} and ground-based observations of
the microlensing event OGLE-2005-SMC-001 to derive the first
microlens parallax measurement using this technique.

\section{Observations
\label{sec:obs}}

On 2005 July 9 (HJD$'\equiv$ HJD-2450000 = 3561.37), the
OGLE-III Early Warning System (EWS, \citealt{udalski03})
alerted the astronomical community that
OGLE-2005-SMC-001 ($\alpha_{{\rm J}2000.0}=0^{\rm h}40^{\rm m}28.\!\!^{\rm s}5$,
$\delta_{{\rm J}2000.0}=-73\grad 44^\prime 46.1^{\prime\prime}$) 
was a probable microlensing event, approximately 23 days
(and seven observations) into the 2005 OGLE-III observing season
for the SMC. In fact, the EWS issued an internal
alert five days earlier, when there were only three 2005 points, but
the OGLE team reacted cautiously because of the high rate of
questionable alerts toward the SMC and because the source lies
projected against a background galaxy, making it a potential
supernova candidate.  However, the event shows a modest, but
unambiguous rise 140 days earlier, at the end of the 2004 season,
which is inconsistent with a supernova, and the unmagnified source
sits right in the middle of the red giant clump on the color-magnitude
diagram (CMD), with $(V-I,I)=(0.92,18.4)$.  
Moreover, the light curve is achromatic.  These factors convinced us that
this was genuine microlensing, leading us to exercise our {\it Spitzer}
ToO option, 
which consisted of three 2-hr observations: two placed symmetrically 
around the peak and one at the baseline.  Once this decision was made, 
OGLE increased its density of coverage to 
3--5 $I$-band observations per clear night.
OGLE observations were obtained using the 1.3 m Warsaw telescope
at Las Campanas Observatory in Chile, operated by the Carnegie
Institution of Washington.  
The photometry was reduced using the standard OGLE-III data pipeline
\citep{udalski03} based on the image subtraction technique, DIA \citep{wozniak}.
Also many $V$-band observations were obtained during the event for 
monitoring achromaticity.
There are {\it Spitzer} observations at four epochs (not three, as
originally envisaged).  These were centered at 2005 July 15 UT
20:02:40, 2005 August 25 UT 12:44:25, 2005 September 15 UT 20:13:53, and 2005
November 29 UT 10:24:40.  The first, third, and fourth observations each 
lasted 2 hr
and consisted of two sets of about 100 dithered exposures, 
each of 26.8 s.  The
second observation (in August) was 1 hr, consisting of one set of
99 dithered
exposures, each of 26.8 s. 
It was obtained with director's
discretionary time (DDT).  All four were carried out simultaneously at
3.6 and 5.8 $\mu$m.  However, the third observation (which took
place at relatively high magnification) was supplemented by 30 minutes
of very short exposures in all four Infrared Array Camera (IRAC) 
filters to probe the detailed
spectral energy distribution of the source.  The reason for the additional
DDT observation is discussed in detail in \S~\ref{sec:complications}.

As originally conceived, the experiment was to consist only of OGLE and
{\it Spitzer} observations.  However, unexpected complications led us
to take additional data from 
other ground-based observatories as well
as the {\it Hubble Space Telescope} (HST).

Initially, we obtained some data using the 1.3 m SMARTS (former Two
Micron All Sky Survey [2MASS])
telescope at the Cerro Tololo Inter-American Observatory (CTIO) in Chile simply
as a precaution against possible future problems with the OGLE telescope.
(In order to align different light curves, it is generally necessary
that they have some overlap; one cannot wait for the problems to
arise before beginning to take data.)  However, as the event approached
peak, we found that it could not be fit with a classical \citet{pac86}
model, even when augmented by parallax.  We therefore began to
intensively observe the event from both the OGLE and SMARTS telescopes 
in the hopes of obtaining
enough data to determine the nature of the light curve anomaly.  Similar
considerations led us to begin observations using the 0.35m Nustrini
telescope at the Auckland Observatory in New Zealand,  which lies at
a substantially  different longitude and suffers from substantially
different weather patterns from those experienced by the two Chile telescopes.

Additionally, several high-dispersion spectra were obtained at Las
Campanas Observatory using 6.5 m Magellan and 2.5 m du Pont telescopes
with echelle spectrographs at different magnifications in order to check 
for potential radial-velocity variations in the spectra of the magnified source.

Finally, the anomalous behavior made it prudent to get high-resolution
images using {\it HST}, both to improve the modeling of blended light
in the ground-based and {\it Spitzer} images, and to determine whether
the apparent ground-based source could actually be resolved into multiple
sources.  We had {\it HST} ToO time to complement low-resolution
space-based microlensing parallax
observations, which was originally to be applied to observations
by {\it Deep Impact}.  With the probability low that these would be triggered
as originally planned, we applied this time to 
obtain two orbits of observations of OGLE-2005-SMC-001.
There were two epochs of $(V,I,J,H,K)$ exposures, on 2005 October 1 and
2006 May 17/18, with exposure times of (300, 200, 351, 351, and 639) seconds.
The infrared observations were then repeated on 2006 June 25.

\subsection{Error Rescaling
\label{sec:error}}

Errors from each ground-based observatory are rescaled to force
$\chi^2$ per degree of freedom close to unity.  For the OGLE data,
we find by inspecting the cumulative distribution of the normalized
residuals $(\delta/\sigma)^2$, that the rescaling factor is not
uniform over the data set.  Here $\sigma$ is the error reported by
OGLE and $\delta$ is the deviation of the data from the model.  We
therefore rescale in 4 segments, which are separated at
HJD = 2453100, 2453576, and 2453609, with rescaling 
factors 1.4, 1.9, 3.6, and 2.0.
We tested two other error-rescaling schemes, one with no rescaling and the
other with uniform rescaling of the OGLE data.  We found that the
solutions do not differ qualitatively when these alternate schemes are
used.

\subsection{{\it Spitzer} Data Reduction and Error Determination
\label{sec:reduction}}
Our scientific goals critically depend on obtaining high-precision
IRAC photometry for each of the 4 epochs.  See \S~\ref{sec:addobs}.
These 4 epochs are divided into 7 1-hour sub-epochs, each consisting
of about 100 dithers, one sub-epoch for the second epoch and two for
each of the other three epochs.
Based on photon statistics alone, the best possible precision would be
about $0.2\%$ for the five sub-epochs near peak, and $\sim 0.4\%$ 
for the last two sub-epochs.  However, there are three
interrelated problems that must be overcome to even approach this
potential.  First, the images contain ``stripes'' produced by
nearby bright stars, perhaps AGB stars, which (because the 3
near-peak observations took place over 60 days) appear at several
different rotation angles.  Indeed, we expended considerable
effort repositioning each successive image to avoid having these
stripes come too close to, or actually overlap, the microlensed target,
but they inevitably did overlap some reference stars.  Second, the
microlensed source is blended with a neighboring star within
$1.\hskip-2pt '' 3$, 
which is easily resolved in {\it HST} images and
clearly resolved in OGLE images as well.  Third, this problem is
significantly
complicated by the well-known fact that IRAC 3.6 $\mu$m images are
undersampled.

We apply the procedures of \citet{reach05} to perform aperture
photometry on the Basic Calibrated Data (BCD), which includes
array-location-dependent and ``pixel-phase'' photometric corrections
at the few percent level.  We choose 7 bright and isolated
stars, which we select from the OGLE images.  The
{\it HST} frames are of course even better resolved, but they are too small
to contain a big enough sample of reference stars.
The centroid position of the target-star aperture on each of the (roughly
100)
BCD dithers is determined by aligning the comparison stars with the OGLE
coordinates.  We determine the ``internal error'' for the target star
and the comparison stars at each sub-epoch from the internal scatters in
their measurements.  This is typically very close to the photon limit.  
However, we find that the epoch-to-epoch scatter in the comparison
stars is about 0.7\%.  While in principle this could be due to intrinsic
variability, such variability is unlikely to be so pervasive at this
level,
particularly since any star showing variability in the $I$ band over several
years was excluded as a reference star.
Hence, we attribute this variation to unknown epoch-to-epoch systematics,
and we assume that these affect the target in the same way that they
affect the reference stars.  Hence, we adopt 0.7\% as our photometric
error for each of the 7 {\it Spitzer} sub-epochs.

We also attempted to do point-spread-function (rather than aperture)
photometry, making use of ``Point Response Functions'' available at 
the {\it Spitzer} Science Center Web site.  However, we
found that the reference stars showed greater scatter between sub-epochs
with this approach and so adopted the results from aperture photometry.

\section{Complications Alter Strategy and Analysis
\label{sec:complications}}

\begin{figure}
\plotone{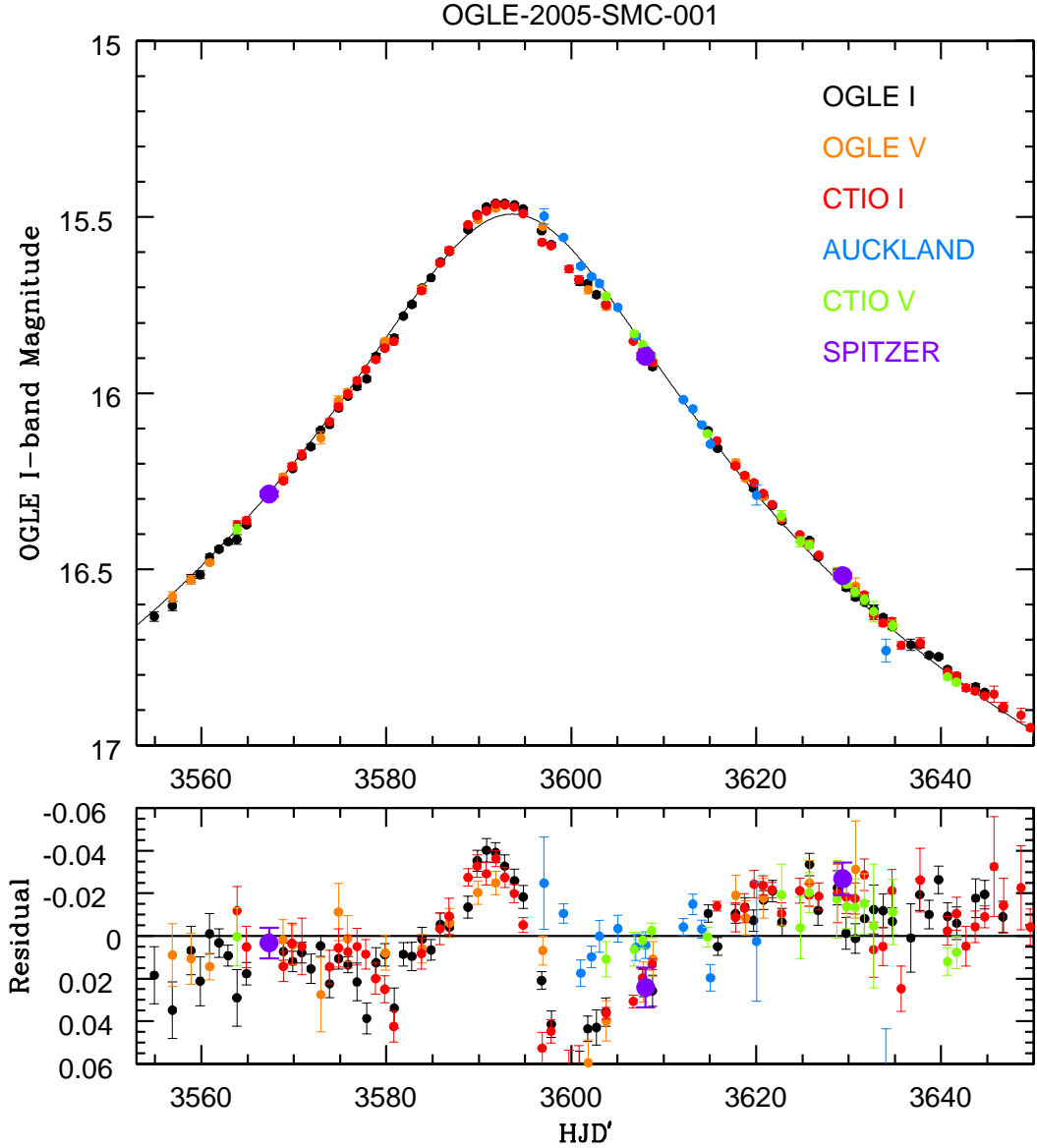}
\caption{\label{fig:pspl}
Standard \citep{pac86} microlensing fit to the light curve of 
OGLE-2005-SMC-001, with data from OGLE $I$ and $V$ in Chile,
$\mu$FUN $I$ and $V$
in Chile, Auckland clear-filter in New Zealand, 
and the {\it Spitzer} satellite 3.6 $\mu$m
at $\sim 0.2\,\au$ from Earth.  The data are binned by the day.
All data are photometrically aligned with the (approximately calibrated)
OGLE data.
The residuals are
severe indicating that substantial physical effects are not being modeled.
The models do not include parallax, but when parallax is included, the
resulting figure is essentially identical.
}\end{figure}

Figure \ref{fig:pspl} shows the ground-based light curve with a fit
to a standard \citep{pac86} model.  The residuals are severe.  The
model does not include parallax.  However, models that include
parallax are quite similar.  In the period before the peak, we were
constantly refitting the light curve with every new night's data in 
order to be able to predict the time of peak and thus the time of the
second ToO observation (which was supposed to be symmetric around the
peak with the first).  It became increasingly clear that the event
was not standard microlensing, and we began to consider alternate
possibilities, including binary source (also called ``xallarap''),
binary lens, and variable source.  The last was especially alarming
because if the variability were irregular, it would be almost impossible
to model at the high precision required to carry out this experiment,
particularly because the source might vary differently at $I$ and $L$.  
Our concern about xallarap led us to obtain radial-velocity measurements
at several epochs near the peak.  These turned out to be the same
within less than $1\,\kms$, which ruled out xallarap for all but the
most pathologically face-on orbits.
But
regardless of the nature of the anomaly, it could potentially cause
serious problems because much more and better data are required
to accurately model complicated light curves compared to simple ones.
Hence, as described in \S~\ref{sec:obs}, recognition of the anomaly
caused us to significantly intensify our ground-based observations.
Moreover, it also caused us to think more carefully about how we
would extract parallax information from a more complicated light curve,
and this led us to recognize a complication that affects even 
light curves that do not suffer from additional anomalies.

\subsection{Need for Additional {\it Spitzer} Observation
\label{sec:addobs}}

Recall that $\pi_{\e,\tau}$ is derived from the different peak time
$t_0$ as seen from Earth and {\it Spitzer}: if the two {\it Spitzer}
observations are timed so that the fluxes seen at Earth are equal
to each other (one on the rising and one on the falling wing of the
light curve), then the {\it Spitzer} fluxes will nevertheless be
different, the first one being higher if the event peaks at {\it Spitzer}
before the Earth.  However, the two {\it Spitzer} fluxes may differ
not only because {\it Spitzer} is displaced from the Earth along
the direction of lens-source relative motion, but also if it is
displaced in the orthogonal direction by {\it different amounts} at
the two epochs.  \citet{gould99} recognized this possibility, but 
argued that the amplitude of this displacement could be determined
from the measurement of $\pi_{\e,\parallel}$, which is derived from
the ground-based parallax measurement (i.e., from the asymmetry of
the light curve).  Hence, he argued that it would be possible to
correct for this additional offset and still obtain a good measurement
of $\pi_{\e,\tau}$.  Unfortunately, while it is true that the {\it amplitude}
can be so derived, the {\it sign} of this correction is more difficult
to determine.

The problem can be understood by considering the work of \citet{smith03},
who showed that ground-based microlensing parallaxes 
are subject to a two-fold degeneracy, essentially whether the source
passes the lens on the same or opposite side of the Earth compared to the Sun.
Within the geocentric formalism of \citet{gould04}, this amounts to
switching the sign of the impact parameter $u_0$ (which by convention
is normally positive) and leaving all other parameters essentially unchanged.
\citet{smith03} derived this degeneracy under the assumption that
the Earth accelerates uniformly during the course of the event.  This
is a reasonable approximation for short events, but is grossly
incorrect for OGLE-2005-SMC-001, with its timescale of $t_\e\sim 0.5\,$yr.
Nevertheless, this degeneracy can hold remarkably well even for relatively
long events, particularly for $|u_0|\ll 1$.   The sign of the
correction to $\pi_{\e,\tau}$ depends essentially on whether the
{\it absolute value} of the impact parameter as seen from {\it Spitzer}
is higher or lower than that seen from Earth.  While the {\it algebraic}
displacement of {\it Spitzer} along this direction can be predicted
from the ground-based measurement of the parallax asymmetry (just
as \citealt{gould99} argued), its effect
on the absolute value of $u_0$ depends on whether $u_0$ is positive
or negative.

If the parallax is sufficiently large, then the ground-based light curve
alone can determine the sign of $u_0$, and if it is sufficiently small,
the difference between the two solutions is also very small and may
not be significant.  However, for intermediate values of the parallax,
this degeneracy can be important.  To understand how an additional
{\it Spitzer} observation can help, consider an idealized set of four
observations, one at peak, one at baseline, and two symmetrically timed
around peak (as seen from Earth).  Call these fluxes $f_P$, $f_{\rm b}$, $f_-$,
and $f_+$.  
Consider now the ratio $[f_P-f_{\rm b}]/[(f_- + f_+)/2 - f_{\rm b}]$.
If the impact parameter seen from {\it Spitzer} is higher than that from 
Earth, this ratio will be lower for the {\it Spitzer} data than for
the ground-based data.  (Note that blended light, which may be different
for the two sets of observations, cancels out of this expression.)
In practice, we found from simulations that it was not necessary to
have the three observations timed so perfectly.  Hence, it was possible
to plan both the additional DDT observation, as well as the second ToO
observation to occur during regularly scheduled IRAC campaigns, so there
was no 6.5 hour penalty for either observation.  Hence, the net cost
to {\it Spitzer} time was less than would have been the case for
a single precisely timed ToO observation.

\begin{figure}
\plotone{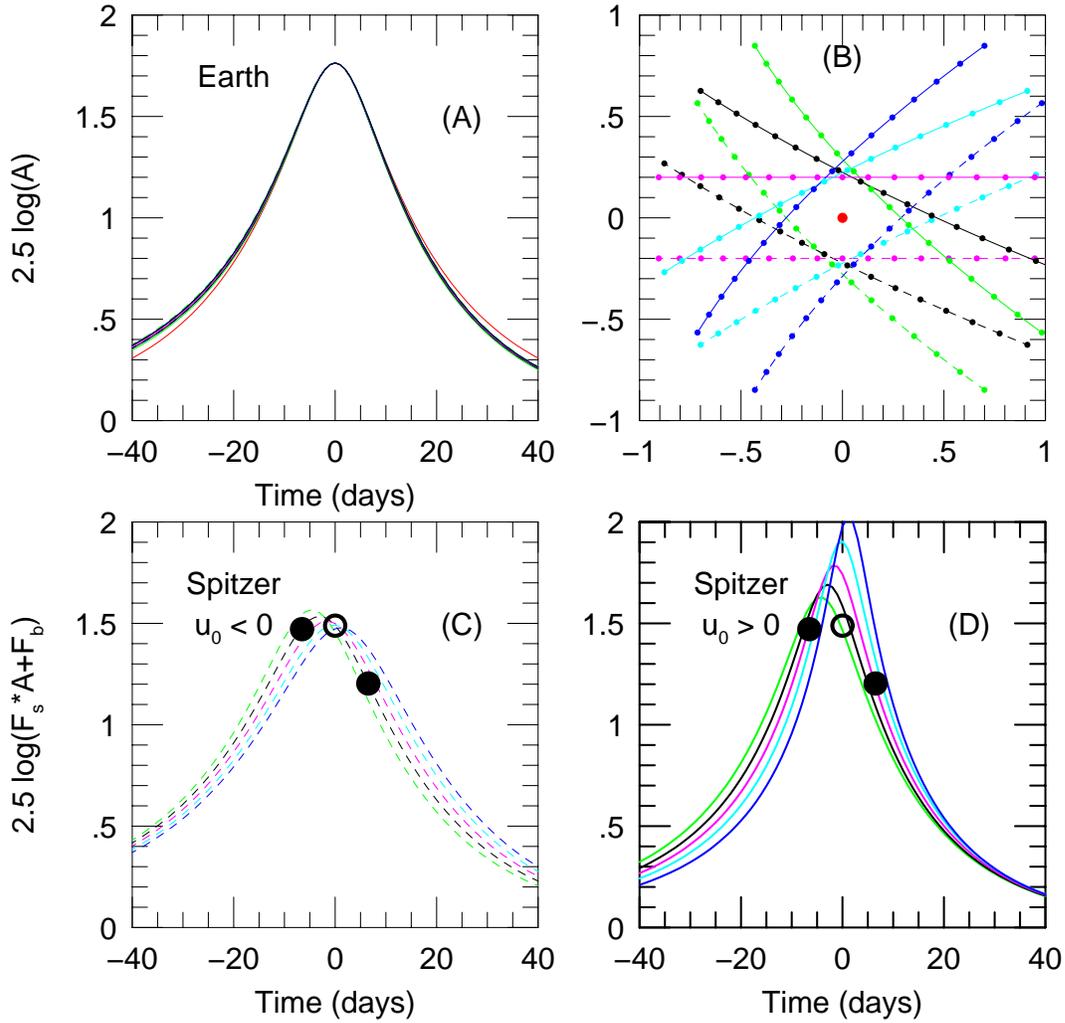}
\caption{\label{fig:degen}
Why 4 (not 3) {\it Spitzer} observations are needed to measure
$\bpi_\e=(\pi_{\e,\parallel},\pi_{\e,\perp})$.  Panel A shows Earth-based
light curve of hypothetical event ({\it black curve}) with 
$\bpi_\e=(0.4,-0.2)$, $u_0=-0.2$, and $t_\e=40\,$days, together with
the corresponding ({\it red}) lightcurve with zero parallax.  From the
asymmetry of the lightcurve, one can measure $\pi_{\e,\parallel}=0.4$
and $|u_0|=0.2$, 
but no information can be extracted about $\pi_{\e,\perp}$ or the
sign of $u_0$.  Indeed, 9 other curves are shown with various values
of these parameters, and all are degenerate with the black curve.
Panel (B) shows the trajectories of all ten models in the geocentric
frame \citep{gould04} that generate these
degenerate curves.  {\it Solid} and {\it dashed} curves indicate
positive and negative $u_0$, respectively, with 
$\pi_{\e,\perp}=-0.4,-0.2,0,+0.2,+0.4$ 
({\it green, black, magenta, cyan, blue}).
Motion is toward positive $x$, while the Sun lies directly toward
negative $x$.  Dots indicate 5 day intervals.
Panel C shows full light curves as would be seen by {\it Spitzer}, located
0.2 AU from the Earth at a projected angle $60^\circ$ from the Sun, for the 5
$u_0<0$ trajectories in Panel B.  The source flux $F_{\rm s}$ and blended flux
$F_{\rm b}$ are fit from the two {\it filled circles} and a third point at
baseline (not shown) as advocated by \citet{gould99}.  Note that these
two points (plus baseline) pick out the ``true'' ({\it black}) trajectory,
from among other solutions that are consistent with the ground-based data
with $u_0<0$, but Panel D shows that these points alone would
pick out the {\it magenta} trajectory among $u_0>0$ solutions, which
has a different $\pi_{\e,\perp}$ from the ``true'' solution.
However, a fourth measurement {\it open circle} would rule out this
{\it magenta} $u_0>0$ curve and so confirm the {\it black} $u_0<0$ curve.
}
\end{figure}

In brief, the above considerations demonstrate that the \citet{gould99}
technique requires a total of four observations, not three as originally
proposed.  Moreover, these observations do not have to be so precisely
timed as \citet{gould99} originally imagined.  
See Figure~\ref{fig:degen}
for a visual explanation of these arguments.

\subsection{Eight-fold Way
\label{sec:binlens}}

Ultimately, we found that the anomaly was caused by a binary lens.
Binary lenses are subject to their own discrete degeneracies.  
This means that analysis of the event is impacted by two distinct
classes of discrete degeneracies: those due to parallax and thoese 
due to binarity.
The discrete parallax degeneracy, as summarized in \S~\ref{sec:addobs}, takes
the impact parameter
$u_0\rightarrow -u_0$ and (because $u_0\ll 1$) leaves other parameters 
changed by very little (see \citealt{gould04}).  

The discrete binary degeneracy is between wide and close binaries.
Here ``wide'' means $b\gg 1$ and ``close'' means $b\ll 1$, where $b$ is the
angular separation between the two binary components in units of $\theta_\e$, 
(for which we give and justify our 
convention in \S~\ref{sec:statparms}).
It gives $(b_c,q_c)\rightarrow (b_w,q_w)$, 
\begin{equation}
b_w = {1+q_c\over 1-q_c}\,b_c^{-1},\qquad
q_w = {q_c\over (1-q_c)^2}
\label{bqtrans}
\end{equation}
and leaves other parameters roughly unchanged.  Here $q$ is the mass ratio
of the lens, with the convention that for $q_w$, the component
closer to the source trajectory goes in the denominator of the ratio.
In both cases, the central magnification pattern
is dominated by a 4-cusp caustic.

This degeneracy was first discovered empirically
by \citet{albrow99} and theoretically by \citet{dominik99} and can
be incredibly severe despite the fact that the two caustics are far
from identical: the solutions can remain indistinguishable
even when there are two well-observed caustic crossings \citep{an05}.  

In the present case, the deviations from a simple lens are not caused
by caustic crossings, but rather by a close approach to a cusp, which
makes this degeneracy even more severe.  In fact,
the caustic is symmetric enough that the approach may almost equally
well be to either of two adjacent cusps.  
That is, the cusp degeneracy would be ``perfect'' if the cusp were
four-fold symmetric, and it is only the deviation from this symmetry
that leads to distinct solutions for different cusp approaches.

In brief, the lens geometry
is subject to an 8-fold discrete degeneracy, 2-fold for parallax,
2-fold for wide/close binary, and 2-fold for different cusp approaches.

\section{Binary Orbital Motion
\label{sec:rotation}}

Of course all binaries are in Kepler orbits, but it is usually possible
to ignore this motion in binary-lens analyses.  Stated less 
positively, it is rarely possible to constrain any binary orbital
parameters from microlensing light curves.  In the few known exceptions,
\citep{albrow00,an02}, the light curve contained several well-measured
caustic crossings that pinned down key times in the trajectory to
$O(10^{-5})$ of an Einstein crossing time.  Hence, we did not expect
to measure binary rotation in the present case in which there are no
such crossings.

We were nevertheless led to investigate rotation by the following
circumstance.  When we initially analyzed the event using only
ground-based data, we found that the best fit (for all 8 discrete
solutions) had negative blended light, roughly $-10\%$ of the source
light, but with large errors and thus consistent with zero at the $1.5\,\sigma$
level.  This was not unexpected.  As mentioned in \S~\ref{sec:intro},
the component of the microlens parallax perpendicular to the Sun,
$\pi_{\e,\perp}$, is generally poorly constrained by ground-based
data alone.  The reason for this is that small changes in $\pi_{\e,\perp}$,
the Einstein timescale $t_\e$, the impact parameter $u_0$, the
source flux $f_{\rm s}$, and the blended flux, $f_{\rm b}$, 
all induce distortions in the
light curve that are symmetric about the peak, and hence all these
parameters are correlated.  Thus, the large errors (and consequent
possible negative values) of $f_{\rm b}$ are just the obverse
of the large errors in $\pi_{\e,\perp}$.  Indeed, this is the reason
for adding in {\it Spitzer} observations.

However, we found that when the {\it Spitzer} observations were added,
the blending errors were indeed reduced, but the actual value
of the blending remained highly negative, near $-10\%$.
This prompted
us to look for other physical effects that could induce distortions in the
light curve that might masquerade as negative blending.  First
among these was binary orbital motion.  Before discussing this motion,
we first review microlensing by static binaries.  

\subsection{Static Binary Lens Parameters
\label{sec:statparms}}

Point-lens microlensing is described by three geometric parameters,
the impact parameter $u_0$ (smallest lens-source angular separation in units
of $\theta_\e$), the time at which the separation reaches this minimum $t_0$,
and the Einstein crossing time $t_\e=\theta_\e/\mu$, where $\mu$ is the
lens-source relative proper motion.

In binary lensing, there are three additional parameters, the projected
separation of the components (in units of $\theta_\e$) $b$, the mass ratio
of the components $q$, and the angle of the source-lens trajectory relative
to the binary axis, $\phi$.  Moreover the first three parameters now require
more precise definition because there is no longer a natural center to
the system.  One must therefore specify where the center of the system is.
Then $u_0$ becomes the closest approach to this center and $t_0$ the
time of this closest approach.  Finally, $t_\e$ is usually taken to
be the time required to cross the Einstein radius defined by the
{\it combined} mass of the two components.  

In fact, while computer programs generally adopt some fairly arbitrary
point (such as the midpoint between the binaries or the binary center of
mass), the symmetries of individual events can make other choices
much more convenient.  That is certainly the case here.  Moreover, symmetry
considerations that are outlined below
will also lead us to adopt a somewhat non-standard $t_\e$
for the wide-binary case, namely the timescale associated with the
mass that is closer to the source trajectory, rather than the total
mass.  To be consistent with this choice, we
also express $b$ as the separation between wide components in units of
the Einstein radius associated with the nearest mass (rather than the
total mass).

As is clear from Figure \ref{fig:pspl}, the light curve is only
a slightly perturbed version of standard (point-lens) microlensing,
which means that it is generated either by the source passing just
outside the central caustic of a close binary (that surrounds both components)
or just outside one of the two caustics of a wide binary (each associated
with one component).  In either case, the standard point-lens parameters
$u_0$ and $t_0$ will be most closely reproduced if the lens center
is placed at the so-called ``center of magnification''.  For close
binaries this lies at the binary center of mass.  For wide binaries,
it lies $q(1+q)^{-1/2}b^{-1}$ from the component that is closer to the 
trajectory.  Hence, it is separated by approximately $bq/(1+q)$ from the
center of mass.  This will be important in deriving 
equation~(\ref{eqn:alphapi}), below.

For light curves passing close to the diamond caustic of a close binary,
the standard Einstein timescale (corresponding to the total mass) will
be very close to the timescale derived from the best-fit point lens
of the same total mass.  However, for wide binaries, the
standard Einstein timescale is longer by a factor $(1+q)^{1/2}$, where
$q$ is the ratio of companion (whether heavier or lighter) to the component
that is approached most closely.  This is because
the magnification is basically due just to this latter component
(with the companion contributing only minor deviations via its shear),
while the usual Einstein radius is based on the {\it total} mass.
For wide binaries, we therefore adopt an Einstein radius and Einstein
timescale reduced by this same factor.

The advantage of adopting these parameter definitions is that,
being fairly well fixed by the empirical light curve, they are
only weakly correlated with various other parameters, some of which
are relatively poorly determined.

\subsection{Binary Orbital Parameters
\label{sec:orbparms}}

While close and wide binaries can be (and in the present case are)
almost perfectly degenerate in the static case, binary orbital motion
has a radically different effect on their respective light curves.
Note first that while 7 parameters would be required to fully describe
the binary orbital motion, even in the best of cases it has
not proven possible to constrain more than 4 of them \citep{albrow00,an02}.
Two of these have already been mentioned, i.e., $b$ and $q$,
from the static case.  For the two binaries for which additional parameters
have been measured, these have been taken to be a uniform rotation rate 
$\omega$ and a uniform binary-expansion rate $\dot b$.

This choice is appropriate for close binaries because for these, the
center of mass is the same as the center of magnification.  Hence
the primary effects of binary motion are rotation of the magnification
pattern around the center of magnification and the change of the magnification
pattern due to changing separation.  Both of these changes may be
(probably are) nonuniform, but as the light curves are insensitive to
such subtleties, the simplest approximation is uniform motion.

However, the situation is substantially more complicated for wide binaries,
consideration of which leads to a different parameterization.
Recall that the wide-binary center of magnification is not at the center
of mass, and indeed is close to one of the components.  Hence, as the
binary rotates, the center of magnification rotates basically with that
component.  Nominally, the biggest effect of this rotation is the resulting
roughly linear motion of the lens center of magnification relative to the
source.  However, the linear component of this motion, i.e., the first
derivative of the motion at the peak of the event, is {\it already}
subsumed in the source-lens relative motion in the static-binary fit.
The first new piece of information about the binary orbital motion is
the second derivative of this motion, i.e., the acceleration.  Note that
the direction of this acceleration is known: it is along the binary axis.
Moreover, for wide binaries, Kepler's Third Law predicts that the periods 
will typically be much longer than the Einstein timescale, so to a reasonable 
approximation, this direction remains constant during the event.
We designate the acceleration (in Einstein radii per unit time squared)
as $\alpha_b$.

The parameter $\alpha_b$ is related to the distance to the lens in
a relatively straightforward way.  For simplicity, assume for the moment
that the center of magnification is right at the position of the
component that is closer to the source trajectory 
(instead of just near it).  The 
3-dimensional acceleration 
of the
component is $a=GM_2/(b r_\e\csc i)^2$, where $M_2$ is the mass of the 
companion to the closer component,
$r_\e$ is the physical Einstein radius, and $i$ is the angle between
the binary axis and the line of sight.  Hence, 
$\alpha_b = a\sin i/r_\e = (q/b^2) GM_1 \sin^3 i/r_\e^3$,
where $M_1$ is the mass of the closer component.  This can be
simplified with the aid of the following three identities:
1) $4GM_1/c^2 = \tilde r_\e\theta_\e$, 2) $r_\e = D_{\rm L}\theta_\e$,
3) $r_\e/\tilde r_\e = D_{\rm LS}/D_{\rm S}$.  Here $D_{\rm L}$
and $D_{\rm S}$ are the distances to the lens and source, and
$D_{\rm LS} = D_{\rm S} - D_{\rm L}$.
We then find,
\begin{equation}
{\alpha_b\over \pi_\e} = 
{\gamma c^2\sin^3 i\over 4 D_{\rm L}\au}\,{D_{\rm S}^2\over D_{\rm LS}^2},
\label{eqn:alphapi}
\end{equation}
where $\gamma \equiv q/b^2$ is the shear.  Note that the shear determines
the size of the caustic and so is one of the parameters that is
most robustly determined from the light curve.
If we were to take account of the offset between $M_1$ and the center
of magnification, the r.h.s. of equation (\ref{eqn:alphapi}) would
change fractionally by the order of $b^{-2}$.

\subsection{Summary of Parameters
\label{parmsum}}

Thus, the model requires a total of 10 geometrical parameters
in addition to the 8 flux parameters ($f_{\rm s}$ and $f_{\rm b}$ for each of 
the 3 ground-based observatories plus {\it Spitzer}).  
These are the three standard microlensing parameters,
$t_0,u_0,t_\e$ (the time of closest approach, separation at closest approach
in units  of $\theta_\e$, and Einstein timescale), the three additional
static-binary parameters $b,q,\phi$ (the binary separation in units
of $\theta_\e$, the binary mass ratio, and the angle of the source trajectory
relative to the binary axis), the two binary-orbit parameters,
$\dot b$ and either $\omega$ (close) or $\alpha_b$ (wide), and the two 
parallax parameters, 
$\bpi_\e = (\pi_{\e,N},\pi_{\e,E})$, where $N$ and $E$ represent the
North and East directions.  These must be specified for 8 different
classes of solutions.

\section{Search for Solutions
\label{search}}

We combine two techniques to identify all viable models of the observed
microlensing light curve: stepping through parameter space on a grid
(grid-search) and 
Markov Chain Monte Carlo (MCMC) \citep{doran03}.

We begin with the simplest class of binary models, i.e., without parallax
or rotation.  Hence, there are six geometric parameters, 
$t_0$, $u_0$, $t_\e$, $b$, $q$, and $\phi$.  We consider classes
of models with $(b,q,\alphasub)$ held fixed, and vary $(t_0,u_0,t_\e)$
to minimize $\chi^2$.  (Note that for each trial model, $f_{\rm s}$ and $f_{\rm b}$
can be determined algebraically from a linear fit, so their evaluations
are trivial.)\ \ This approach identifies four solutions, i.e. 
(2 cusp-approaches)$\times$(wide/close degeneracy).  We then introduce
parallax, and so step over models with  
$(b,q,\alphasub,\pi_{\e,N},\pi_{\e,E})$ held fixed, working in the neighborhood
of the $(b,q,\alphasub)$ minima found previously.  The introduction of
parallax brings with it the $\pm u_0$ degeneracy, and so there are now
8 classes of solutions.

Next we introduce rotation.  We begin by employing grid search and
find, somewhat surprisingly, that several of the eight
(close/wide, $\pm u_0$, on/off-axis cusp) classes of solutions have
more than one minimum in $(\omega,\dot b)$ for close binaries or
$(\alpha_b,\dot b)$ for wide binaries.  We then use each of these
solutions as seeds for MCMC and find several additional minima that
were too close to other minima to show up in the grid search.
Altogether there are 20 separate minima, 12
for close binaries and 8 for wide binaries.

We use MCMC to localize our solutions accurate to about $1\,\sigma$
and to determine the covariance matrix of the parameters.  In MCMC,
one moves randomly from one point in parameter space to another.
If the $\chi^2$ is lower, the new point is added to the ``chain''.  
If not, one draws a random number and adds the new point only if
this number is lower than relative probability ($\exp(-\Delta\chi^2/2)$).
If the parameters are highly correlated (as they are in microlensing)
and the random trial points are chosen without reference (or without proper
reference) to these covariances, then the overwhelming majority of
trial points are rejected.  
We therefore sample parameter space based on the covariance matrix
drawn from the previous ``links'' in the chain. During the initial
"burning in" stage of the MCMC, we frequently evaluate the covariance
matrix (every 100 ``links'') until it stabilizes. Then we hold the
covariance matrix fixed in the simulation \citep{doran03}.
 From the
standpoint of finding the best $\chi^2$, one can combine linear fits
for the flux parameters $f_{\rm s}$ and $f_{\rm b}$ with MCMC for the remaining
parameters. However, since part of our MCMC objective is to find the 
covariances, we treat $(f_{\rm s},f_{\rm b})_{\rm OGLE}$ as MCMC parameters
while fitting for the remaining four flux parameters analytically.

In order to reduce the correlations among search parameters, we
introduce the following parameter combinations into the search:
$t_{\rm eff}\equiv u_0 t_\e$, $f_{\rm base} = f_{\rm s} + f_{\rm b}$,
$f_{\rm max}\equiv f_{\rm s}/u_0$, and $\gamma\equiv q/b^2$ (wide)
or $Q\equiv b^2 q/(1+q)^2$ (close).  Because these are directly
related to features in the light curve, they are less prone to
variation than the naive model parameters.  The effective timescale
$t_{\rm eff}$ is $1/\sqrt{12}$ of the full-width at half-maximum,
 $f_{\rm base}$ is just the flux at baseline, and $f_{\rm max}$
is the flux at maximum.  The scale of the \citet{cr1,cr2}
distortion (which governs the binary perturbation) is given by
the shear $\gamma$ for wide binaries and by the quadrupole
$Q$ for close binaries.

The MCMC ``chain'' automatically samples points in the neighborhood of the
minimum with probability density proportional to their likelihoods,
$\exp(-\chi^2/2)$.  Somewhat paradoxically, this means that for 
higher-dimensional problems, it does not actually get very close to
the minimum.  Specifically, for a chain of length $N$ sampling an
$m$-dimensional space, there will be only one point 
for which $\Delta\chi^2$ (relative to the minimum) obeys
$\Delta\chi^2<Y$, where $P[\chi^2(m\ {\rm dof})<Y]=N^{-1}$.  Hence, for
$m\gg 1$ one requires a chain of length 
$N\sim e^{1/2}(m/2)!2^{m/2}$ to reach $1\,\sigma$ above
the true minimum, or $N\sim 10^{4.9}$ for $m=12$.  Further
improvements scale only as $\sim N^{1/m}$.  Hence, to find the
true minimum, we construct chains 
in which the rejection criterion is calculated based on
$\exp(-25\chi^2/2)$ rather than $\exp(-\chi^2/2)$.  
However, when calculating error bars
and covariances, or when integrating over the MCMC, we use the
$\exp(-\chi^2/2)$ chain.

\subsection{Convergence
\label{sec:converge}}

A general problem in MCMC fitting is to determine how well the solution
``converges'', that is, how precisely the ``best fit'' solution is
reproduced when the initial seed solution is changed.  In our case,
the problem is the opposite: MCMC clearly does not converge to a
single minimum, and the challenge is to find all the local minima.
We described our procedure for meeting this challenge above, first
by identifying 8 distinct regions of parameter space semi-analytically,
and then exploring these with different MCMC seeds.  This procedure
led to well-defined minima (albeit a plethora of them), whose individual
structures were examined by putting boundaries into the MCMC code that 
prevented the chain from ``drifting'' into other minima.  We halted
our subdivision of parameter space when the structure on the $\chi^2$
surface fell to of order $\Delta\chi^2\sim 1$, regarding the
$\Delta\chi^2\la 1$ region as the zone of convergence.  As mentioned
above, we located the final minimum by artificially decreasing the
errors by a factor of 5, again making certain that the resulting
(exaggerated) $\chi^2$ surface was well behaved.

\section{Solution Triage
\label{sec:triage}}

Table 1 gives parameter values and errors for a total of 20 different
discrete solutions, which are labeled by (C/W) for close/wide binary,
($+$/$-$) for the sign of $u_0$, ($\parallel$/$\perp$) for solutions that
are approximately parallel or perpendicular to the binary axis,
and then by alphabetical sequential 
for different viable combinations of rotation
parameters.  In Table 1, we allow a free fit to blending.

Note that some solutions have severe negative blending.  While 
it is possible in principle that these are due to systematic errors, the
fact that other solutions have near-zero blending and low $\chi^2$
implies that the negative-blending solutions probably have lens
geometries that do not correspond to the actual lens.  Other solutions
have relatively high $\chi^2$ and so are also unlikely.  Solutions
with $\chi^2$ that
is less than 9 above the minimum have their $\chi^2$ displayed in boldface,
while the remaining solutions are shown in normal type.

There are several reasons to believe that the blending is
close to zero.  First, the source appears isolated on our $K$-band
NICMOS {\it HST} images, implying that if it is blended, this blended
light must be within of order 100 mas of the source.  As the density
of sources in the {\it HST} images is low, this is a priori very 
unlikely unless the blended light comes from a companion to the source
or the lens.
Moreover, all the near neighbors of the source on the {\it HST}
image are separately resolved
by the OGLE photometry, so if there is blended light in the OGLE photometry
then it must also be blended in the {\it HST} images.  
Second, the $V-I$ color of the source,
which can be derived by a model-independent regression of $V$ flux
on $I$ flux, is identical within measurement error to the color of
the 
baseline light from the combined source and (possible) blend.
This implies that either 1) the source is unblended,
2) it is blended by another star of nearly the same
color as the source, or 3) it is blended by a star that is so faint
that it hardly contributes to the color of the blend. The second
possibility is strongly circumscribed by the following argument. 
The source is a clump star.  On the SMC CMD, there are first ascent giants 
of color similar to the source from the clump itself down
to the subgiant branch about 2.5 mag
below the clump.  Thus, in principle, the blended light could lie
in the range $0.1\la f_{\rm b}/f_{\rm s}\la 1$ without causing the
baseline color to deviate from the source color.  However, there are 
no solutions in Table 1 with blending this high.  There is one low-$\chi^2$
solution with $f_{\rm b}/f_{\rm s}\sim 0.06$, but this would be 3 mag below
the clump and so below
the subgiant branch.  If we restrict the blend to turnoff colors, i.e.,
$V-I=0.6$, then the color constraint implies 
$f_{\rm b}/f_{\rm s}= 0.01\pm 0.01$, which is negligibly small from our
perspective.  Thus, while it is possible in principle that the source is
blended with a reddish subgiant, the low stellar density in the {\it HST}
image, the low frequency of such subgiants on the CMD, and the difficulty
of matching color constraints even with such a star, combine to make this
a very unlikely possibility.  We therefore conduct the primary analysis
assuming zero blending, as listed in Table 2.  
Again, the $\Delta\chi^2<9$ solutions are marked
in boldface.  Note that most solutions in Table 1 are reasonably
consistent with zero blending, which is the expected behavior for
the true solution provided it is not corrupted by systematic errors.
For these, the $\chi^2$ changes only modestly from Table 1 to Table 2.
However, several solutions simply disappear from Table 2.  This
is because, in some cases, forcing the blending to zero has the 
effect of merging two previously distinct binary-rotation minima.

The main parameters of interests are $\bpi_\e$ and the closely
related quantity $\tilde\bv$.  However, for reasons that will be
explained below, $\tilde\bv$ can be reliably calculated only for
the close solutions, but not the wide solutions.  Hence, we focus
first on $\bpi_\e$.

\begin{figure}
\plotone{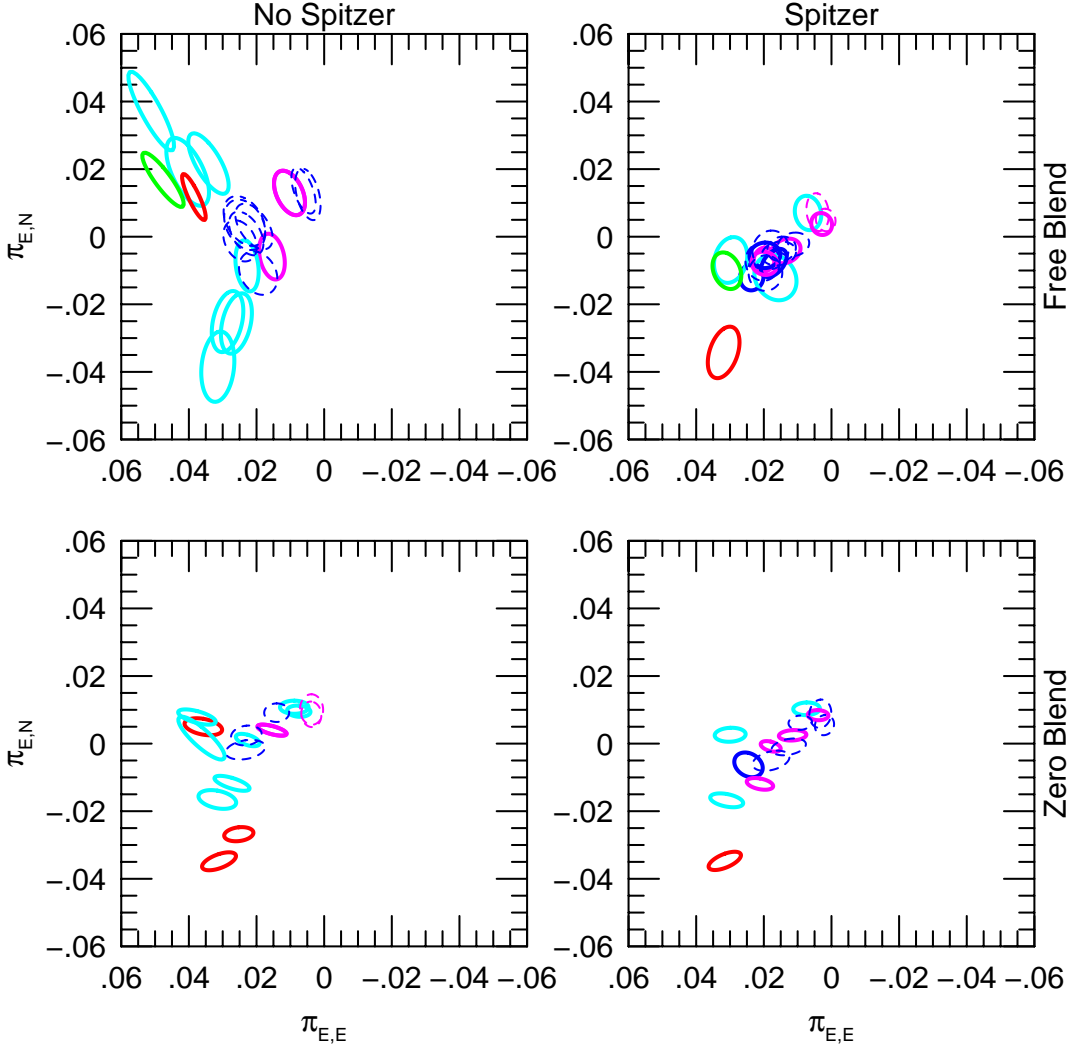}
\caption{\label{fig:pie}
Parallax $\bpi_\e=(\pi_{\e,N},\pi_{\e,E})$ $1\,\sigma$ error ellipses 
for all discrete solutions for OGLE-2005-SMC-001.  The left-hand panels
show fits excluding the {\it Spitzer} data, while the right-hand panels
include these data.  The upper panels show fits with blending as a free
parameter whereas the lower panels fix the OGLE blending at zero.
The ellipses are coded by $\Delta\chi^2$ (relative to each global minimum),
with $\Delta\chi^2<1$ (red), $1<\Delta\chi^2<4$ (green),
$4<\Delta\chi^2<9$ (cyan), $9<\Delta\chi^2<16$ (magenta),
$\Delta\chi^2>16$ (blue).  Close- and wide-binary solutions are represented
by bold and dashed curves, respectively.  Most of the 
``free-blend, no-Spitzer'' solutions are highly degenerate along the
$\pi_{\e,\perp}$ direction ($33^\circ$ north through east), as predicted from
theory, because only the orthogonal ($\pi_{\e,\parallel}$) direction is
well constrained from ground-based data.
At seen from the two upper panels, the {\it Spitzer} 
observations reduce the errors in the $\pi_{\e,\perp}$ direction
by a factor $\sim 3$ when the blending is a free parameter.  However,
fixing the blending (lower panels) already removes this freedom, so
{\it Spitzer} observations then have only a modest additional effect.
}\end{figure}

\begin{figure}
\plotone{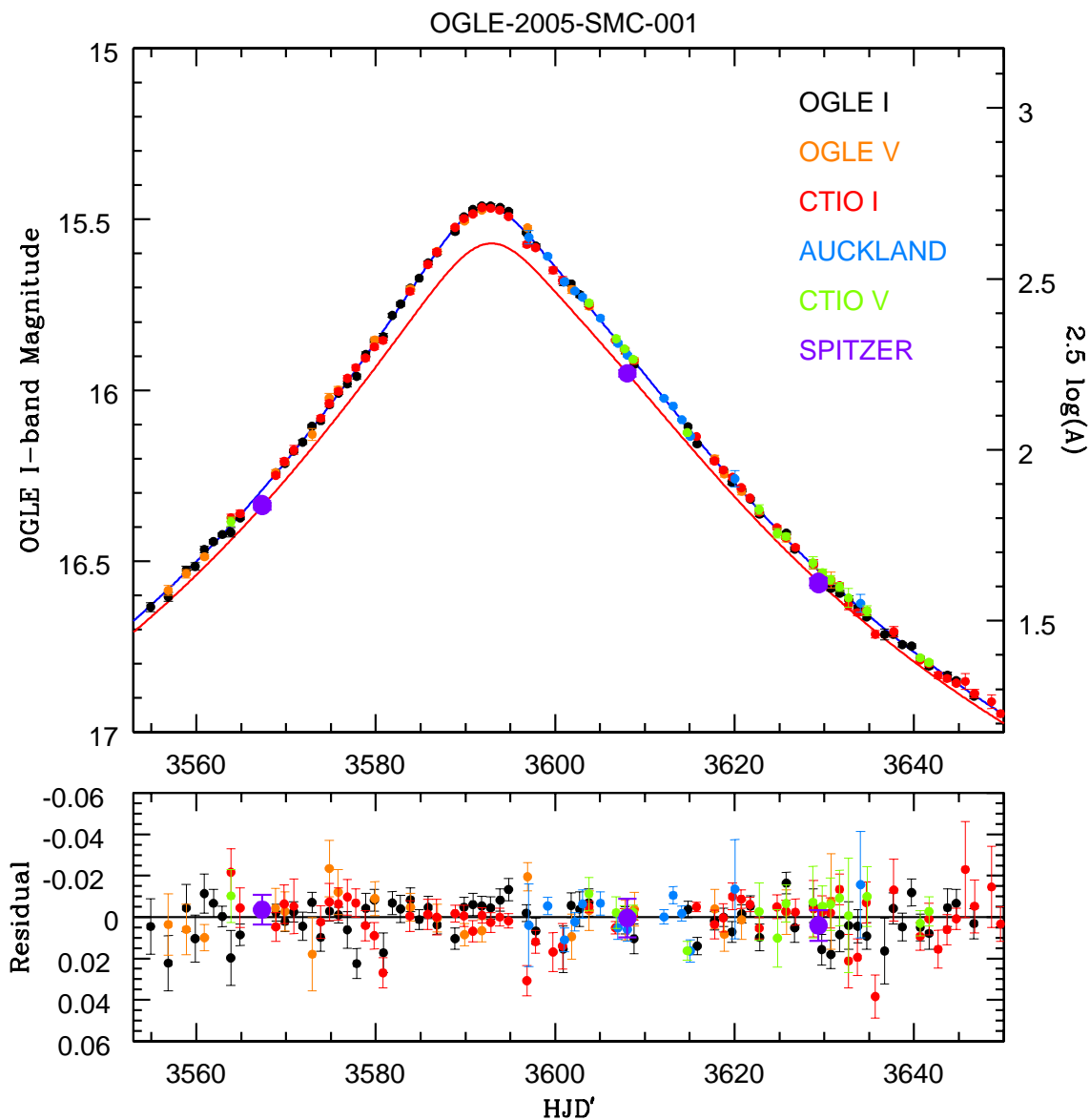}
\caption{\label{fig:bestfit}
Best-fit binary microlensing model for OGLE-2005-SMC-001 together
with the same data shown in Fig.~\ref{fig:pspl}. The model includes 
microlens parallax (two parameters) and binary rotation (two parameters). 
The models for ground-based and {\it Spitzer} observations are plotted 
in blue and red, respectively. All data are in the units of 2.5 log(A), 
where A is the magnification. Ground-based data are also photometrically 
aligned with the (approximately calibrated) OGLE data. The residuals show 
no major systematic trends.
}\end{figure}

Figure~\ref{fig:pie} shows error ellipses for all solutions, color-coded
by according to $\Delta\chi^2$ relative to the global minimum.
The right-hand panels show the solutions presented in Tables 1 and 2, which
include the {\it Spitzer} data.  The left-hand panels exclude these
data.  The upper panels are based on a free fit for blending whereas
the lower panels are constrained to zero blending for the OGLE dataset.
Comparing the two upper panels, it is clear that when blending is a 
fitted parameter, the {\it Spitzer} data reduce the errors in the
$\pi_{\e,\perp}$ direction by about a factor of 3.  However, once
the blending is fixed (lower panels), the {\it Spitzer} data have only
a modest additional effect.  This is expected since $\pi_{\e,\perp}$ is
correlated with blending and one can simultaneously constrain both 
parameters either by constraining $\pi_{\e,\perp}$ with {\it Spitzer}
data or just by fixing the blending by hand.  
Figure \ref{fig:bestfit}
shows the best overall zero-blending fit to the data.

\subsection{Wide-Binary Solutions
\label{sec:wide}}

All eight wide solutions are effectively excluded.  When a free fit
to blending is allowed, their $\chi^2$ values are already significantly
above the minimum.  When zero blending is imposed, only five independent
solutions survive and those that have
negative blending are driven still higher.   In all five
cases to $\Delta\chi^2>16$.

\subsection{Close-Binary Solutions
\label{sec:close}}

\begin{figure}
\plotone{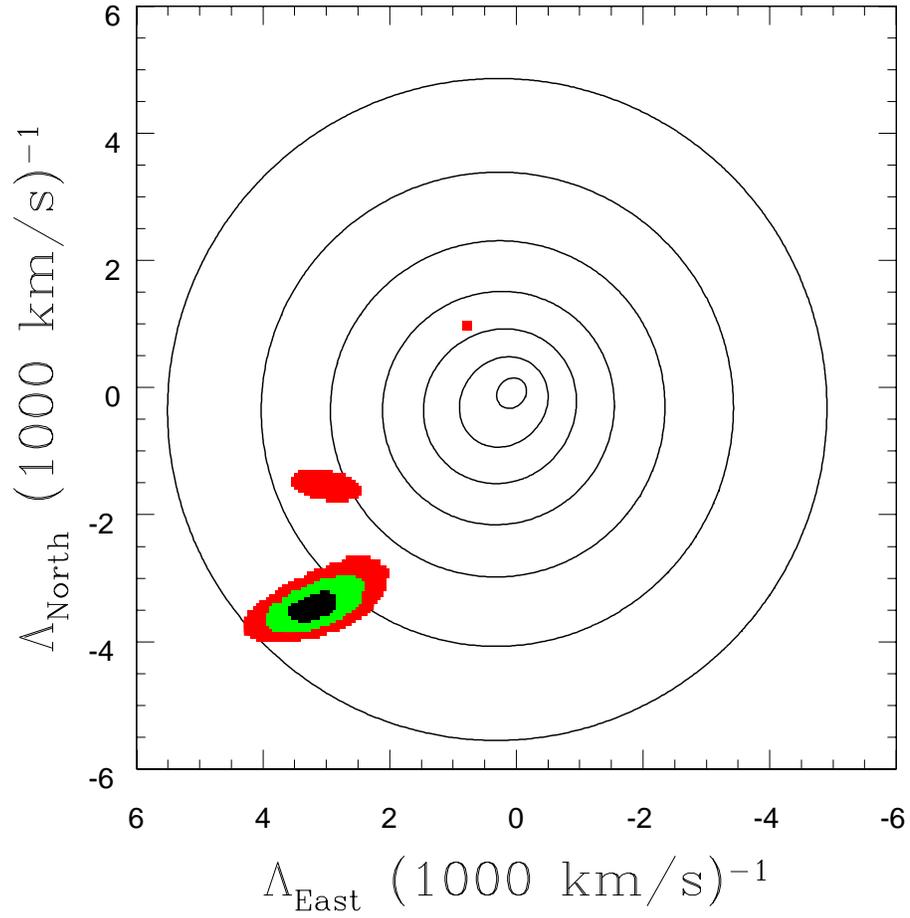}
\caption{\label{fig:lambdasmc}
Likelihood contours of the inverse projected velocity
$\bLambda\equiv \tilde \bv/\tilde v^2$ for SMC lenses together
with $\bLambda$ values for light-curve solutions found by
MCMC.  The latter are color-coded for solutions
with $\Delta\chi^2$ within 1, 4, and 9 of the global minimum.
The likelihood contours are spaced by factors of 5.
}\end{figure}

\begin{figure}
\plotone{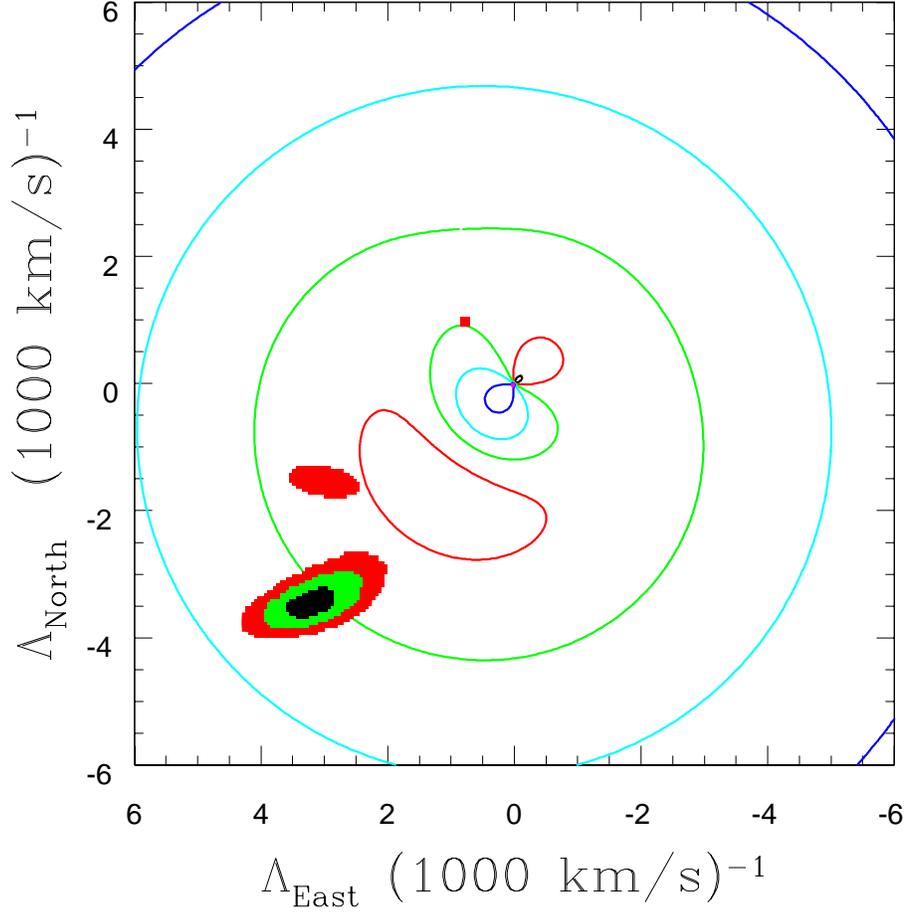}
\caption{\label{fig:lambdahalo}
Likelihood contours of the inverse projected velocity
$\bLambda\equiv \tilde \bv/\tilde v^2$ for halo lenses together
with $\bLambda$ values for light-curve solutions found by
MCMC.  Similar to Figure~\ref{fig:lambdasmc}
except in this case the contours are color coded, with
black, red, yellow, green, cyan, blue, magenta, going from
highest to lowest.
}\end{figure}

Eight Of the 11 close solutions survive the imposition of zero blending.
Of these, only one has $\Delta\chi^2<4$, and only another 
two have $\Delta\chi^2<8.7$
relative to the best zero-blending solution.
We focus primarily on these 
three, which all have
best-fit parallaxes in the range $0.030<\pi_\e<0.047$ and projected
velocities in the range $210\,\kms < \tilde v < 330\,\kms$.  These
projected velocities are of the order expected for halo lenses but
are about 1 order of magnitude smaller than those expected for
SMC self lensing.  Note that because there are multiple solutions,
the errors in $\tilde {\bf v}$ are highly non-Gaussian and are best
judged directly from Figures \ref{fig:lambdasmc} and 
\ref{fig:lambdahalo} (below) rather than quoting a formal error bar.

\section{Lens Location
\label{sec:location}}

When \citet{alcock95} made the first 
measurement of microlensing parallax,
they developed a purely kinematic method of estimating the lens 
distance (and thus mass) based on comparison of the measured value of 
$\tilde \bv$ with the expected kinematic properties of the underlying
lens population.  Starting from this same approach,
\citet{assef06} devised a test that uses the microlens parallax measurement
to assign relative probabilities to different lens populations
(e.g., SMC, Galactic halo, Galactic disk) based solely on the 
{\it kinematic} characteristics of these populations, and without
making prior assumptions about either the mass function or the 
density normalization of any population.  This is especially useful
because, while a plausible guess can be given for the mass function and
normalization of SMC lenses, nothing is securely known about a
putative Galactic halo population.  In the present case, the high
projected velocity $\tilde v$
immediately rules out Galactic-disk lenses, so we 
restrict consideration to the other two possibilities.  

We begin by recapitulating the \citet{assef06} test in somewhat more
general form.  The differential 
rate of microlensing events of fixed mass $M$ (per steradian) is
\begin{equation}
d^{(6)}\Gamma(M)\equiv {d^6\Gamma(M)\over d^2 v_\l\, d^2 v_\s d D_\l d D_\s} =
f_\l(\bv_\l) f_\s(\bv_\s)D_\s^2 \nu_\s(D_\s)\nu_\l(D_\l)2\tilde v\tilde r_\e
{D_\ls^2\over D_\s^2},
\label{eqn:difgam}
\end{equation}
where $f_L(\bv_L)$ and $f_S(\bv_S)$ are the two-dimensional 
normalized velocity distributions of the lenses and sources, 
$\nu_\l$ and $\nu_\s$ are the density distributions of the lenses and
sources, $\tilde v$ is an implicit function of 
$(D_\l,D_\s,\bv_\l,\bv_\s)$, and $\tilde r_\e$ is an implicit function of 
$(D_\l,D_\s,M)$.
The method is simply to evaluate the likelihood
\begin{equation}
{\cal L} = 
{\int d^2 v_\l\, d^2 v_\s d D_\l d D_\s\, d^{(6)}\Gamma(M)
\exp[-\Delta\chi^2(\tilde \bv)/2]
\over\int d^2 v_\l\, d^2 v_\s d D_\l d D_\s\, d^{(6)}\Gamma(M)}
\label{eqn:likerat}
\end{equation}
for each population separately, and then take the ratio of likelihoods
for the two populations: ${\cal L}_{\rm ratio}\equiv
{\cal L}_{\rm halo}/{\cal L}_{\rm SMC}$.
Here, $\Delta\chi^2(\tilde \bv)$ is the difference of $\chi^2$ relative
to the global minimum that is derived from the microlensing light curve.
Note that all dependence on $M$ disappears from ${\cal L}$.
Equation (\ref{eqn:likerat}) can be simplified in different ways for each
population.  In both cases, we express the result in terms of
$\bLambda$,
\begin{equation}
\bLambda \equiv {\tilde \bv\over {\tilde v}^2} = {\bpi_\e t_\e\over\au},
\label{eqn:blamdadef}
\end{equation}
rather than $\tilde \bv$, because it is better behaved in the 
neighborhood of $\bLambda=0$,
just as trigonometric parallax is better behaved near zero than
its inverse, distance. 

\subsection{Halo Lenses
\label{sec:halolens}}

For halo lenses, the depth of the SMC is small compared to $D_\ls$
and the internal dispersion of SMC sources is small compared to the
bulk motion of the SMC.  Hence, one can essentially drop the three
integrations over SMC sources, implying
\begin{equation}
\tilde \bv = {D_\s\over D_\ls}\bv_\l - \bv_\oplus - {D_\l\over D_\ls}\bv_{\rm SMC},
\label{eqn:vleq}
\end{equation}
where $\bv_\l$, $\bv_\oplus$, and $\bv_{\rm SMC}$ are the velocities of the
lens, the ``geocentric frame'', and the SMC (all in the Galactic frame) 
projected on the plane of the sky.  
The ``geocentric frame'' is the frame of the Earth at the time of the peak
of the event. It is the most convenient frame for analyzing microlensing
parallax \citep{gould04} and for this event is offset from the heliocentric
frame by
\begin{equation}
 (v_{\odot,\rm N},v_{\odot,\rm E})-(v_{\oplus,\rm N},v_{\oplus,\rm E})
= (-24.9,-15.5)\,\kms.
\label{eqn:vrel}
\end{equation}

We assume an isotropic Gaussian
velocity dispersion for the lenses with $\sigma_{\rm halo}= 
v_{\rm rot}/\sqrt{2}$,
where $v_{\rm rot}= 220\,\kms$.  After some manipulations (and dropping
constants that would cancel out between the numerator and denominator)
we obtain
\begin{equation}
{\cal L}_{\rm halo} = 
{\int \exp[-\Delta\chi^2(\bLambda)/2]
g_{\rm halo}(\bLambda,D_\l) d D_\l d\Lambda_{\rm North}d\Lambda_{\rm East}\over
\int g_{\rm halo}(\bLambda,D_\l) d D_\l d\Lambda_{\rm North}d\Lambda_{\rm East}},
\label{eqn:likehalo}
\end{equation}
where
\begin{equation}
g_{\rm halo}(\bLambda,D_\l) = \exp(-v_\l^2/2\sigma_{\rm halo}^2)\nu_{\rm halo}
(D_\l)\Lambda^{-5}
D_\ls^{7/2} D_\l^{1/2},
\label{eqn:ghalo}
\end{equation}
and where $v_\l$ is an implicit function of $\bLambda$ through
equations (\ref{eqn:blamdadef}) and (\ref{eqn:vleq}).  We adopt
$\nu_{\rm halo}({\bf r}) = const/(a_{\rm halo}^2 + r^2)$, where 
$a_{\rm halo}=5\,\kpc$
and the Galactocentric distance is $R_0=7.6\,\kpc$.
We adopt $\bv_\odot = (10.1,224,6.7)\,\kms$ in Galactic coordinates,
which leads to a 2-dimensional projected velocity of 
$(v_{\odot,\rm N},v_{\odot,\rm E})= (126,-126)\,\kms$ toward the SMC source.
From our assumed distances $D_{\rm SMC}=60\,\kpc$ and the SMC's measured
proper motion of $(-1.17\pm 0.18,+1.16\pm 0.18)$
\citep{kallivayalil06}, we obtain,
\begin{equation}
 (v_{\rm SMC,\rm N},v_{\rm SMC,\rm E})-(v_{\odot,\rm N},v_{\odot,\rm E})
= (-333,330)\,\kms.
\label{eqn:vrel2}
\end{equation}

\subsection{SMC Lenses
\label{sec:smclens}}

For the SMC, we begin by writing 
\begin{equation}
\tilde \bv = {\bv_\l/D_\l - \bv_\s/D_\s\over D_\l^{-1}-D_\s^{-1}}
=\bv_\s + {D_\s\over D_\ls}\Delta\bv
\rightarrow\bv_{\rm SMC} - \bv_\oplus + {D_{\rm SMC}\over D_\ls}\Delta\bv,
\label{eqn:vtildesmc}
\end{equation}
where $\bv_\l$ and $\bv_\s$ are now measured in the geocentric frame and
$\Delta \bv\equiv \bv_\l- \bv_\s$.  The last step in
equation~(\ref{eqn:vtildesmc}) is an appropriate
approximation because the SMC velocity dispersion is
small compared to its bulk velocity and $D_\ls\ll D_\s$.   

We now assume that the sources and lenses are drawn from the same population,
which implies that the dispersion of $\Delta\bv$ is larger than those
of $\bv_\l$ and $\bv_\s$ by $2^{1/2}$.  We assume that this
is isotropic with Gaussian dispersion $\sigma_{\rm SMC}$.  Again making the
approximation $D_\s\rightarrow D_{\rm SMC}$, we can factor the
integrals in equation~(\ref{eqn:likerat}) by evaluating the density
integral
\begin{equation}
\eta(D_{\rm LS}) = \int d D_\l \nu(D_\l)\nu(D_\l+D_\ls),
\label{eqn:etadef}
\end{equation}
where $\nu=\nu_\l=\nu_\s$.
We then obtain
\begin{equation}
{\cal L}_{\rm SMC} = 
{\int \exp[-\Delta\chi^2(\bLambda)/2] 
g_{\rm SMC}(\bLambda,D_\ls) d D_\ls d\Lambda_{\rm North}d\Lambda_{\rm East}\over
\int g_{\rm SMC}(\bLambda,D_\ls) d D_\ls 
d\Lambda_{\rm North}d\Lambda_{\rm East}},
\label{eqn:likesmc}
\end{equation}
where
\begin{equation}
g_{\rm SMC}(\bLambda,D_\l) = \exp[-(\Delta v)^2/4\sigma_{\rm SMC}^2]
\eta(D_\ls)D_\ls^{7/2}\Lambda^{-5},
\label{eqn:gsmc}
\end{equation}
and where $\Delta v$ is an implicit function of $\bLambda$ through
equations (\ref{eqn:blamdadef}) and (\ref{eqn:vtildesmc}).  

\subsection{SMC Structure
\label{sec:smcstruct}}

In order to evaluate equation~(\ref{eqn:gsmc}) one must estimate
$\sigma_{\rm SMC}$ as well as
the SMC density $\nu$ along the line of sight, which is required
to compute $\eta(D_\ls)$.  This requires an investigation of the
structure of the SMC.

In sharp contrast to its classic ``Magellanic irregular''
appearance in blue light, the SMC is essentially
a dwarf elliptical galaxy whose old population is quite regular in
both its density \citep{zaritsky00} and velocity \citep{harris06}
distributions.  \citet{harris06} find that after removing an
overall gradient (more below), the observed radial velocity distribution
is well fit by a Gaussian with $\sigma =27.5\,\kms$,
which does not vary significantly over their $4^\circ\times 2^\circ$
(RA,Dec) field.

We adopt the following SMC parameters for the 1-dimensional dispersion 
$\sigma_{\rm SMC}$, the tidal radius $r_t$, and the density profile
(along the line of sight), $\nu(r)= const/(a_{\rm SMC}^2 + r^2)^{n/2}$,
\begin{equation}
\sigma_{\rm SMC} = 25.5\,\kms, 
\quad r_t = 6.8\,\kpc,
\quad a_{\rm SMC} = 1\,\kpc,
\quad n = 3.3,
\label{eqn:smcparms}
\end{equation}
as we now justify.

The $\sigma =27.5\,\kms$ dispersion reported by \citet{harris06} includes
measurement errors.  When the reported errors (typically $10\,\kms$ per star)
are included in the fit, this is reduced to $\sigma_{\rm SMC} = 24.5\,\kms$.
However, these reported errors may well be too generous: the statistical
errors (provided by D.~Zaritsky 2006, private communication) are typically only
2--3$\,\kms$, the reported errors being augmented to account for
systematic errors.  If the statistical errors are used, we find
$\sigma_{\rm SMC} = 26.5\,\kms$.  D.~Zaritsky (2006, private communication)
advocates an intermediate value for this purpose, which leads to 
$\sigma_{\rm SMC} = 25.5\,\kms$.

The old stellar population in the SMC is rotating at most very slowly.
\citet{harris06} report a gradient across the SMC of $8.3\,\kmsd$, which
they note is a combination of the traverse velocity of the SMC and
the solid-body component of internal bulk motion.  As the SMC proper motion
was poorly determined at the time, \citet{harris06} did not attempt
to disentangle these two.  However, \cite{kallivayalil06} have now measured
the SMC proper motion to be  $(\mu_{\rm N},\mu_{\rm E})= 
(-1.16\pm 0.18,1.17\pm 0.18)\,\masyr$.  We refit the \citet{harris06} data
and find 
$\nabla v_r = (-10.5\pm 1.4,5.0\pm 0.7)\,\kmsd$.  Subtracting these two 
measurements (including errors and covariances) we obtain a net internal rotation
of $5.2\pm 1.6\,\kmsd$ with a position angle of $183^\circ\pm 30^\circ$.
Since this rotation is due north-south (within errors), a direction
for which the data have a baseline of only $\sim \pm 1^\circ$, and
since solid-body rotation is unlikely to extend much beyond the core,
it appears that the amplitude of rotational motion is only about
$5\,\kms$, which is very small compared to the dispersion, $\sigma_{\rm SMC}$.
Hence, we ignore it.  In addition, we note that this rotation is
misaligned with the HI rotation axis \citep{stanimirovic04}
by about $120^\circ$,
so its modest statistical significance may indicate that it is not real.

As we describe below, our likelihood estimates are fairly
sensitive to the tidal radius $r_t$ of the SMC.
Proper determination of the tidal radius is a complex problem.
Early studies, made before dark matter was commonly accepted, were
carried out for Kepler potentials and in analogy with stellar and
solar-system problems (e.g.~\citealt{king62}).  \citet{read06} have
calculated tidal radii for a range of potentials and also for an
orbital parameter $\alpha$ that ranges from $-1$ for retrograde to
$+1$ for prograde.  We choose $\alpha=0$ as representative and evaluate
their expression for an isothermal potential and for the satellite
being close to pericenter (as is appropriate for the SMC):
\begin{equation}
{r_t\over D} = {\sigma_{\rm sat}\over\sigma_{\rm host}}
\biggl(1+ {2\ln\xi\over1-\xi^{-2}}\biggr)^{-1/2}.
\label{eqn:tidal}
\end{equation}
Here, $\xi$ is the ratio of the apocenter to the pericenter of the
satellite orbit, $D$ is the pericenter distance,
and $\sigma_{\rm sat}$ and $\sigma_{\rm host}$ are the respective
{\it halo} velocity dispersions.  Because we adopt an $n=3.3$ profile,
the SMC halo velocity dispersion is larger than its stellar
dispersion by $(3.3/2)^{1/2}$, implying that
${\sigma_{\rm sat}/\sigma_{\rm host}}= 3.3^{1/2}\sigma_{\rm SMC}/v_{\rm rot}
=0.21$.  We adopt $\xi=3$ based on typical orbits found by
\citet{kallivayalil06}, which yields $r_t = 0.107\,D_{\rm SMC} = 6.8\,\kpc$.
Note, moreover, that at $\xi=3$, $d\ln r_t/d\ln \xi\sim 0.24$, so the
tidal radius is not very sensitive to the assumed properties of the
orbit.

The most critical input to the likelihood calculation is the stellar
density along the line of sight.  Of course, images of the SMC give direct
information only about its surface density as a function of position.
One important clue to how the two are related comes from the HI velocity
map of \citet{stanimirovic04}, which shows an inclined rotating
disk with the receding side at a position angle of $\sim 60^\circ$ (north
through east).  This is similar to the $\sim 48^\circ$
position angle of the old-star optical profile found by \citet{harris06}
based on data from \citet{zaritsky00}.  Hence, the spheroidal old-stellar
population is closely aligned to the HI disk,
although (as argued above) the stars
in the SMC are pressure- rather than rotationally-supported. 
Our best clue to the line-of-sight profile is the major-axis profile
exterior to the position of the source, which lies about $1\,\kpc$
to the southwest of the Galaxy center, roughly along the apparent major axis.
The projected surface density is falling roughly as $r^{-2.3}$ over the
$\sim 1\,\kpc$ beyond the source position, from which we derive a
deprojected exponent of $n=3.3$.  This is similar to the exponent for
Milky Way halo stars.  While there is a clear core in the star counts, this
may be affected by crowding, and the core seems to have little impact
on the counts beyond the source position (which is what is relevant to
the density profile along the line of sight).  Hence, we adopt 
$a_{\rm SMC}=1\,\kpc$.

The likelihood ratio is most sensitive to the assumptions made about the
stellar density in the outskirts of the SMC, hence to the power law
and tidal radius adopted.  This seems strange at first sight because
the densities in these outlying 
regions
 are certainly extremely small,
whatever their exact values.  This apparent paradox can be understood
as follows.  For $D_\ls\la r_t$, the leading term in $\eta$ is
$\eta(D_\ls)\sim [D_\ls^2 + 4(b^2 + a_{\rm SMC}^2)]^{-n/2}$,
where $b=1.0\,\kpc$ is the impact parameter.  
In the outskirts, this implies $\eta(D_\ls)\sim D_\ls^{-n}$. 
The integrand in the numerator of equation~(\ref{eqn:likesmc})
then scales as $\exp\{-[\Delta v(\bLambda,D_\ls)]^2/4\sigma_{\rm SMC}^2\} 
D_\ls^{7/2-n}$.  For fixed $\bLambda$, $\Delta v$ is
a rising function of $D_\ls$, and so for sufficiently large $D_\ls$,
the exponential will eventually cut off the integral.  However, for
the measured value of $\bLambda$ (corresponding to 
$\Delta \tilde v\equiv 
|\bLambda/\Lambda^2-(\bv_{\rm SMC}-\bv_\odot)|\sim 300\,\kms$,
the cutoff does not occur until 
$D_\ls \sim (2\sigma_{\rm SMC}/\tilde v)D_{\rm SMC}\sim 10\,\kpc$.
Thus, as long as the density exponent remains $n\la 3.5$ and as long
as the density is not actually cut off by $r_t$, the integral keeps
growing despite the very low density.  On the other hand, a parallel
analysis shows that the denominator in equation~(\ref{eqn:likesmc})
is quite insensitive to assumptions about the outer parts of the SMC.

\subsection{Likelihood Ratios
\label{sec:likerats}}

Using our adopted SMC parameters (eq.~[\ref{eqn:smcparms}]), we find 
${\cal L}_{\rm ratio}\equiv
{\cal L}_{\rm halo}/{\cal L}_{\rm SMC}=27.4$.  As discussed in
\S~\ref{sec:smcstruct}, this result is most sensitive to the
outer SMC density profile, set by the exponent $n$ and the cutoff
$r_t$.  If the tidal radius is increased from $r_t=6.8\,\kpc$ to 
$r_t=10\,\kpc$ and other parameters are held fixed,
${\cal L}_{\rm ratio} = 30.3$.  If the exponent is reduced from $n=3.3$
to $n=2.7$, then ${\cal L}_{\rm ratio} = 29.1$.  Hence, halo lensing
is strongly favored in any case, but
by an amount that would vary noticeably if any of our key model
parameters were markedly off.

In carrying out these evaluations, we integrated equations 
(\ref{eqn:likehalo}) and (\ref{eqn:likesmc}) over all solutions
with $\Delta\chi^2<9$.  These solutions are grouped around three close-binary
minima shown in Table 2.  We carried out the integration in two different
ways: over the discrete ensemble of solutions found by the Markov chains
and uniformly over the three error ellipses that were fit to these chains.
The results do not differ significantly.  Figures \ref{fig:lambdasmc}
and \ref{fig:lambdahalo} show $\bLambda$ for 
the Markov-chain solutions superposed on
likelihood contours for SMC and halo lenses respectively.

\subsection{Kepler Constraints for Close Binaries
\label{sec:closeconstraints}}

Binaries move in Kepler orbits.  In principle, if we could measure
all the Kepler parameters then these, together with the measured microlens
parallax $\bpi_\e$ (and the approximately known source distance $D_\s$), 
would fix the mass and distance to the lens.  While the two orbital parameters
that we measure are not sufficient to determine the lens mass and distance,
they do permit us to put constraints on these quantities.

Consider first the special case of a face-on binary in a circular orbit.
Kepler's Third Law implies that $GM/(b r_\e)^3 = \omega^2$ where
$M$ is the mass of the lens, $b$ is the binary separation in units of the
local Einstein radius $r_\e$, and $\omega$ is the measured rotation 
parameter.  Since $r_\e=(D_\ls/D_\s)\tilde r_\e$, this implies,
\begin{equation}
{c^2\over 4 \tilde r_\e b^3}\,{D_\s^2\over D_\l D_\ls^2}
=\omega^2\qquad \rm (face-on\ circular)
\label{eqn:faceoncirc}
\end{equation}
If one considers other orientations but remains restricted to face-on
orbits, then this equation becomes a (``greater than'') inequality
because at fixed projected separation,
the apparent angular speed can only decrease.

Further relaxing to the case of non-circular orbits but with $\dot b/b=0$
(i.e., the event takes place at pericenter), the rhs of 
equation~(\ref{eqn:faceoncirc}) is halved, $\omega^2\rightarrow \omega^2/2$,
because escape speed (appropriate for near-parabolic orbits)
is $\sqrt{2}$ larger than circular speed.  Finally, after some algebra,
and again working in the parabolic limit, one finds that including
non-zero radial motion leads to 
$\omega^2/2\rightarrow [\omega^2+ (\dot b/b)^2/4]/2$, and thus to
\begin{equation}
{D_\l D_\ls^2\over D_\s^2} \leq {c^2\over 2\tilde r_\e b^3}
\biggl[\omega^2 + {(\dot b/b)^2\over 4}\biggr]^{-1}.
\label{eqn:genorbit}
\end{equation}

If the rhs of this equation is sufficiently small, then only lenses
that are near the Sun ($D_\l\ll D_\s$) or near the SMC ($D_\ls\ll D_\s$)
will satisfy it.  In practice, we find that this places no constraint
on SMC lenses, but does restrict halo lenses to be relatively close
to the Sun, with the limit varying from $\sim 2\,\kpc$ to $\sim 10\,\kpc$
depending on the particular solution being probed and the MCMC realization 
of that solution.  Integrating over the entire Markov chain, we find
that ${\cal L}_{\rm ratio}$ is reduced by a factor three, from 27.4 to 11.2.

\subsection{Kepler Constraints for Wide Binaries
\label{sec:wideconstraints}}

As discussed in \S~\ref{sec:wide}, all wide solutions are ruled out
by their high $\chi^2$.
Nevertheless, for completeness 
it is instructive to ask what sort of constraints could
be put on the lens from measurement of the acceleration parameter $\alpha_b$
if these solutions had been accepted as viable.  We rewrite 
equation~(\ref{eqn:alphapi}) in terms of the parameter combination
$T\equiv (\pi_\e\gamma/\alpha_b)^{1/2}$, which has units of time,
\begin{equation}
\sqrt{D_\l\over 60\,\kpc}
{D_{\rm LS}\over D_{\rm S}} = {cT\over 2}\sqrt{\sin^3 i\over 60\,\kpc\,\au}
< 0.28\,{T\over \rm yr}.
\label{eqn:zeval}
\end{equation}

The five wide-binary zero-blending
solutions listed in Table 2 have, respectively,
$T=(0.039,0.032,0.035,0.017,0.015)$ yr.  Thus, if
these solutions had been viable, the lens would have been firmly 
located in the SMC (or else, improbably, within 6 pc of
the Sun).  This demonstrates the power of this constraint
for wide binaries, which derives from light-curve features that
arise from the motion of the center of magnification relative to the
center of mass.  These obviously do not apply to close binaries
for which the center of magnification is identical to the center of mass.

However, this same relative motion makes it {\it more difficult} to
constrain the nature of the lens from measurements of the projected
velocity, $\tilde \bv$.   This is because the parallax measurement
directly yields only the projected velocity of the center of magnification,
whereas models of the lens populations constrain the motion of the
center of mass.  If $\dot b$ is measured, it is straight forward
to determine the component of the difference between these two
that is parallel to the binary axis, namely
\begin{equation}
 \Delta \tilde v_\parallel = {\dot b\over 1+q}\,{\au\over\pi_\e}.
\label{eqn:tildevpar}
\end{equation}

On the other hand, the measurement of $\alpha_b$ indicates that there
is motion in the transverse direction, but specifies neither the
amplitude nor sign.  For example, for circular motion with the
line of nodes perpendicular to the binary axis,
\begin{equation}
 \Delta \tilde v_\perp = \pm \sqrt{\alpha_b b\over 1+q}\,{\au\over\pi_\e}
\csc i,
\label{eqn:tildevperp}
\end{equation}
and hence,
\begin{equation}
 {\Delta \tilde v_\perp\over\tilde v} = 
\pm \sqrt{\alpha_b b\over 1+q}\,t_\e \csc i,
\label{eqn:tildevperp2}
\end{equation}
If this ratio (modulo the $\csc i$ term) is small, then the internal
motion can be ignored (assuming the inclination is not unluckily low).
In the present case, the quantity 
$[\alpha_b b/(1+q)]^{1/2}t_\e$ for the five respective wide-binary
solutions shown in Table 2 is (0.23,0.36,0.25,1.04,0.77), implying
that constraints arising from measurement of $\tilde v$ would be significantly
weakened.  

Again, however, since the wide solutions are in fact ruled out, all of these
results are of interest only for purposes of illustration.

\subsection{Constraints From (Lack of) Finite-Source Effects
\label{sec:finitesource}}

Finite source effects are parameterized by $\rho\equiv\theta_*/\theta_\e$,
where $\theta_*$ is the angular size of the source.  Equivalently,
$\rho = (D_\l/D_\ls)(r_*/\tilde r_\e)$, where $r_*=D_\s\theta_*$ is the
physical source radius.  Since the source is a clump giant, with 
$r_*\sim 0.05\,\au$, this implies 
\begin{equation}
\rho = 0.021{D_L\over 60\,\kpc}\,{5\,\kpc\over D_\ls}\,{\pi_\e\over 0.035}
\label{eqn:rhoeq}
\end{equation}
All models described above assume $\rho=0$.
Since the impact parameter is $u_0\sim 0.08$ and the semi-diameter of
the caustic is $2Q\sim 0.02$, finite source effects should be pronounced
for $\rho\ga u_0-2Q\sim 0.06$.  In fact, we find
that zero-blending models with $\rho\la0.05$ are contraindicated by
$\Delta\chi^2\sim (\rho/0.0196)^4$.
This further militates against SMC lenses, which
predict acceptable values of $\rho$ only for relatively large
source-lens separations.  If we penalize solutions with finite
source effects by $\Delta\chi^2=(\rho/0.0196)^2$, we find that
${\cal L}_{\rm ratio}$ rises from 11.2 to 20.3.

\section{Discussion
\label{sec:discussion}}

By combining ground-based and {\it Spitzer} data, we have measured
the microlensing parallax accurate to 0.003 units, by far the best
parallax measurement yet for an event seen toward the Magellanic Clouds.
Our analysis significantly favors a halo location for the lens over
SMC self lensing.  It excludes altogether lensing by Galactic disk
stars.  Of course, with only one event analyzed using this technique,
an SMC location cannot be absolutely excluded based on the $\sim 5\%$
probability that we have derived.  The
technique must be applied to more events before firm conclusions
can be drawn.  {\it Spitzer} itself could be applied to this task.
Even better would be
observations by the {\it Space Interferometry Mission (SIM)}
\citep{gs99}, for which time is already allocated for 5 Magellanic
Cloud events.  {\it SIM} would measure both $\bpi_\e$ and $\theta_\e$ and
so determine (rather than statistically constrain) the position of the
lens.

Assuming that the lens is in the halo, what are its likely properties?
The mass is 
\begin{equation}
M=10.0\,M_\odot\biggl({\pi_\e\over 0.047}\biggr)^{-2}\,
{\pi_\rel\over 180\,\muas},
\label{eqn:halomass}
\end{equation}
where the fiducial $\pi_\e$ is the best-fit value and the fiducial $\pi_\rel$
is for a ``typical'' halo lens, which (after taking account of the
constraint developed in \S~\ref{sec:closeconstraints}) would lie at 
about 5 kpc.  In the best-fit model, the
mass ratio is $q=2.77$, which would imply primary and secondary masses
of 7.3 and 2.7 $M_\odot$ respectively.  The projected separation would be
$b=0.22$ Einstein radii, i.e., $4.7\,\au(D_\ls/D_\s)$. 
Note that at these relatively close distances, main-sequence stars
in this mass range would shine far too brightly to be compatible
with the strict constraints on blended light.  Hence, the lenses
would have to be black holes.

We must emphasize that the test carried out here {\it uses only kinematic
attributes} of the lens populations and
assumes no prior information about {\it their mass functions}.
Note, in particular, that if we were to adopt as a priori the lens mass 
distribution inferred by the MACHO experiment \citep{alcock00}, then
the MACHO hypothesis would be strongly excluded.  Recall from 
\S~\ref{sec:closeconstraints} (eq.~[\ref{eqn:genorbit}]), that the lens
is either very close to the SMC or within 10 kpc of the Sun.  However,
from equation (\ref{eqn:halomass}), for $D_{\rm L}<10\,\kpc$, we have
$M\ga 4.6\,M_\odot$, and hence a primary mass $Mq/(1+q)\ga 3.4\,M_\odot$.
 From Figures 12--14 of \citet{alcock00}, such masses are strongly 
excluded as the generators of the microlensing events they observed toward
the LMC.

Hence, if the MACHO hypothesis favored by this single event is
correct, the MACHO population must have substantially different
characteristics from those inferred by \citet{alcock00}, in particular,
as mentioned above, a mass scale of order $10\,M_\odot$ or perhaps more.
\citet{alcock01b} limit the halo fraction of such objects to $<30\%$
at $M=10\,M_\odot$ and to $<100\%$ at $M=30\,M_\odot$.  At higher
masses, the sensitivity of microlensing surveys deteriorates drastically.
However, \citet{yoo04} derived important limits in this mass range
from the distribution of wide binaries in the stellar halo, putting
an upper limit at 100\% for $M=40\,M_\odot$ and at 20\% for
$M>200\,M_\odot$.  Thus, if the MACHO hypothesis is ultimately
confirmed, this would be a new population in the mass ``window'' 
identified in Figure 7 of \citet{yoo04} between the limits set by microlensing
and wide-binary surveys.  This again argues for the importance of
obtaining space-based parallaxes on additional microlensing events.

As we discussed in some detail in \S~\ref{sec:smcstruct} (particularly
the last paragraph) our conclusion regarding the relatively low
probability of the lens being in the SMC rests critically on the
assumption that the SMC lens population falls off relatively
rapidly along the line of sight.  (We adopted an $n=3.3$ power law.)\ \
This is because equation (\ref{eqn:gsmc}) scales $\propto D_\ls^{7/2-n}$.
If the SMC had a halo lens population with a much shallower falloff than we
have assumed,
this term could dominate the integral in equation (\ref{eqn:likesmc})
even if the overall normalization of the halo were relatively low.
Hence, Magellanic Cloud halo lensing could provide an alternate
explanation both for this SMC event and the events seen by MACHO toward
the LMC \citep{calchi06}.  As mentioned above, {\it SIM} could easily 
distinguish between this conjecture and the halo-lens hypothesis.

Finally, we remark that analysis of this event was extraordinarily
difficult because it was a ``weak'' (i.e., non-caustic-crossing) binary.
If it had been either a single-mass lens or a caustic-crossing binary,
it would have been much easier to analyze and the inferences regarding
its location (in the halo or the SMC) would have been much more
transparent.  We therefore look forward to applying this same technique
to more typical events.

\acknowledgments
We thank Scott Gaudi for a careful reading of the manuscript, which
led to many useful suggestions. We are also grateful to Patrick 
Tisserand for his comments. This work is based in part on
observations made with the {\it Spitzer Space Telescope}, which
is operated by the Jet Propulsion Laboratory, California
Institute of Technology, under a contract with NASA. Support
for this work was provided by NASA through contract 1277721
issued by JPL/Caltech. Support for program \#10544 was provided 
by NASA through a grant from the Space Telescope Science Institute, 
which is operated by the Association of Universities for Research in 
Astronomy, Inc., under NASA contract NAS 5-26555.
S.D. and A.G. were supported in part by grant AST 042758 from the NSF.
Support for OGLE was provided by Polish MNiSW grant  N20303032/4275, NSF
grant AST-0607070 and NASA grant NNG06GE27G.
Any opinions, findings, and conclusions or recommendations expressed in
this material are those of the authors and do not necessarily reflect the
views of the NSF. We thank the Ohio Supercomputer Center for the use 
of a Cluster Ohio Beowulf cluster in conducting this research.

\clearpage

\begin{deluxetable}{lrrrrrrrrrrr}                                                                                                           
\tabletypesize{\footnotesize}
\tablewidth{520pt}
\tablecaption{\label{tab:modelfree} Light Curve Models: Free Blending.  }                                                       
\vskip 1em                                                                                                              
\startdata
Model & {$t_0$} & {$u_0$} & {$t_{\rm E}$} & {$q$} & {$Q:\gamma$} & {$\phi$} & {$\pi_{{\rm E},N}$} & {$\pi_{{\rm E},E}$} 
& {$\omega:\alpha_b^{1/2}$} & {$\dot b/b$} & {$F_{\rm b}/F_{\rm base}$}\\                                               
$\chi^2$    & (day) &$(\times 100)$& (day) &  &$(\times 100)$& (deg) & & & (yr$^{-1}$)& (yr$^{-1}$) & \\ \hline         
\hline                                                                                                                  
C-$\parallel $a&$3593.751$&$-8.729$&$174.17$&$ 2.77$&$0.980$&$ 21.70$&$-0.0342$&$ 0.0319$&$ 0.073$&$ 5.60$&$ 0.003$\cr
  {\bf 1455.40}&$   0.040$&$ 0.532$&$  8.13$&$ 0.33$&$0.066$&$  1.10$&$ 0.0076$&$ 0.0047$&$ 0.047$&$  0.46$&$ 0.061$\cr
 \hline
C+$\parallel $a&$3593.687$&$ 7.763$&$190.03$&$ 2.15$&$0.850$&$ 19.34$&$-0.0120$&$ 0.0236$&$-0.141$&$ 5.83$&$ 0.115$\cr
        1475.74&$   0.033$&$ 0.351$&$  6.58$&$ 0.26$&$0.046$&$  1.26$&$ 0.0039$&$ 0.0035$&$ 0.027$&$  0.48$&$ 0.040$\cr
 \hline
C-$\parallel $b&$3593.560$&$-9.567$&$160.61$&$ 0.98$&$0.999$&$  9.23$&$-0.0069$&$ 0.0298$&$ 0.331$&$ 0.40$&$-0.101$\cr
  {\bf 1462.89}&$   0.036$&$ 0.509$&$  6.69$&$ 0.25$&$0.063$&$  1.89$&$ 0.0068$&$ 0.0049$&$ 0.032$&$  0.61$&$ 0.060$\cr
 \hline
C+$\parallel $b&$3593.643$&$ 9.321$&$163.41$&$ 1.11$&$1.002$&$ 11.24$&$-0.0068$&$ 0.0199$&$-0.167$&$ 1.33$&$-0.071$\cr
        1466.84&$   0.040$&$ 0.346$&$  4.52$&$ 0.15$&$0.046$&$  1.12$&$ 0.0036$&$ 0.0032$&$ 0.029$&$  0.26$&$ 0.040$\cr
 \hline
C-$\parallel $c&$3593.473$&$-9.020$&$171.06$&$ 0.29$&$0.837$&$ -3.34$&$ 0.0070$&$ 0.0069$&$ 0.746$&$-3.87$&$-0.030$\cr
  {\bf 1463.52}&$   0.042$&$ 0.352$&$  5.89$&$ 0.04$&$0.041$&$  1.31$&$ 0.0051$&$ 0.0039$&$ 0.069$&$  0.44$&$ 0.043$\cr
 \hline
C+$\parallel $c&$3593.460$&$ 9.207$&$167.46$&$ 0.29$&$0.853$&$ -3.78$&$ 0.0038$&$ 0.0029$&$-0.743$&$-3.92$&$-0.053$\cr
        1465.15&$   0.042$&$ 0.276$&$  4.15$&$ 0.03$&$0.035$&$  1.24$&$ 0.0032$&$ 0.0031$&$ 0.070$&$  0.41$&$ 0.033$\cr
 \hline
C-$\parallel $d&$3593.639$&$-9.831$&$157.29$&$ 0.90$&$1.052$&$  8.53$&$-0.0121$&$ 0.0166$&$ 0.146$&$-0.04$&$-0.131$\cr
  {\bf 1464.13}&$   0.058$&$ 0.480$&$  6.18$&$ 0.15$&$0.062$&$  1.18$&$ 0.0067$&$ 0.0062$&$ 0.096$&$  0.27$&$ 0.057$\cr
 \hline
C+$\parallel $d&$3593.676$&$ 9.406$&$162.74$&$ 0.79$&$1.001$&$  7.86$&$-0.0042$&$ 0.0135$&$-0.144$&$-0.12$&$-0.080$\cr
        1467.70&$   0.056$&$ 0.333$&$  4.42$&$ 0.14$&$0.043$&$  1.47$&$ 0.0036$&$ 0.0043$&$ 0.048$&$  0.50$&$ 0.040$\cr
 \hline
C+$\parallel $e&$3593.655$&$ 9.161$&$165.60$&$ 1.10$&$0.983$&$ 11.36$&$-0.0055$&$ 0.0197$&$-0.162$&$ 1.50$&$-0.051$\cr
        1474.32&$   0.047$&$ 0.498$&$  7.24$&$ 0.20$&$0.060$&$  1.42$&$ 0.0038$&$ 0.0045$&$ 0.038$&$  0.74$&$ 0.058$\cr
 \hline
C+$\parallel $f&$3593.769$&$ 7.516$&$194.33$&$ 1.86$&$0.826$&$ 19.21$&$-0.0078$&$ 0.0170$&$-0.018$&$ 5.94$&$ 0.140$\cr
        1474.40&$   0.038$&$ 0.372$&$  7.10$&$ 0.33$&$0.048$&$  1.46$&$ 0.0042$&$ 0.0038$&$ 0.041$&$  0.72$&$ 0.042$\cr
 \hline
C-$\perp     $a&$3593.251$&$-8.026$&$176.25$&$ 1.74$&$0.719$&$280.21$&$-0.0101$&$ 0.0309$&$ 0.758$&$ 1.10$&$ 0.061$\cr
  {\bf 1457.55}&$   0.030$&$ 0.370$&$  6.94$&$ 0.20$&$0.041$&$  0.53$&$ 0.0054$&$ 0.0043$&$ 0.049$&$  0.34$&$ 0.046$\cr
 \hline
C+$\perp     $a&$3593.272$&$ 7.900$&$177.32$&$ 1.40$&$0.752$&$280.03$&$-0.0079$&$ 0.0198$&$-0.599$&$ 1.45$&$ 0.073$\cr
        1470.74&$   0.035$&$ 0.298$&$  5.33$&$ 0.25$&$0.052$&$  0.58$&$ 0.0035$&$ 0.0038$&$ 0.084$&$  0.32$&$ 0.037$\cr
 \hline
W-$\parallel $a&$3593.708$&$-8.470$&$170.22$&$ 2.49$&$0.936$&$  6.38$&$ 0.0073$&$ 0.0041$&$ 0.241$&$ 0.49$&$-0.004$\cr
        1471.32&$   0.037$&$ 0.306$&$  5.70$&$ 0.40$&$0.036$&$  0.51$&$ 0.0056$&$ 0.0033$&$ 0.175$&$  0.02$&$ 0.038$\cr
 \hline
W+$\parallel $a&$3593.703$&$ 8.568$&$167.72$&$ 2.48$&$0.950$&$  6.37$&$ 0.0050$&$ 0.0017$&$ 0.273$&$ 0.49$&$-0.016$\cr
        1471.29&$   0.035$&$ 0.273$&$  4.70$&$ 0.40$&$0.034$&$  0.49$&$ 0.0033$&$ 0.0029$&$ 0.174$&$  0.02$&$ 0.033$\cr
 \hline
W-$\perp     $a&$3593.562$&$-8.666$&$174.41$&$ 2.28$&$1.154$&$279.74$&$-0.0039$&$ 0.0180$&$ 0.841$&$-0.75$&$ 0.009$\cr
        1474.79&$   0.067$&$ 0.361$&$  6.31$&$ 1.03$&$0.048$&$  0.61$&$ 0.0057$&$ 0.0054$&$ 0.385$&$  0.06$&$ 0.043$\cr
 \hline
W+$\perp     $a&$3593.641$&$ 9.281$&$165.12$&$ 7.39$&$1.158$&$279.74$&$-0.0036$&$ 0.0148$&$ 0.804$&$-0.23$&$-0.060$\cr
        1480.79&$   0.062$&$ 0.293$&$  4.04$&$ 5.71$&$0.048$&$  0.60$&$ 0.0036$&$ 0.0042$&$ 0.339$&$  0.18$&$ 0.034$\cr
 \hline
W-$\perp     $b&$3593.570$&$-9.209$&$166.04$&$ 3.21$&$1.207$&$279.24$&$-0.0074$&$ 0.0204$&$ 0.909$&$-0.16$&$-0.054$\cr
        1475.50&$   0.062$&$ 0.385$&$  5.83$&$ 1.92$&$0.053$&$  0.58$&$ 0.0056$&$ 0.0051$&$ 0.360$&$  0.18$&$ 0.046$\cr
 \hline
W+$\perp     $b&$3593.659$&$ 8.897$&$171.08$&$ 5.28$&$1.125$&$280.32$&$-0.0023$&$ 0.0112$&$ 0.826$&$-0.77$&$-0.016$\cr
        1480.16&$   0.068$&$ 0.255$&$  3.89$&$ 3.53$&$0.042$&$  0.61$&$ 0.0035$&$ 0.0046$&$ 0.305$&$  0.06$&$ 0.030$\cr
 \hline
W-$\perp     $c&$3593.492$&$-9.443$&$163.30$&$ 2.20$&$1.291$&$278.45$&$-0.0108$&$ 0.0195$&$ 1.094$&$ 0.49$&$-0.086$\cr
        1474.90&$   0.046$&$ 0.329$&$  5.24$&$ 0.51$&$0.052$&$  0.48$&$ 0.0053$&$ 0.0050$&$ 0.244$&$  0.05$&$ 0.040$\cr
 \hline
W+$\perp     $c&$3593.534$&$ 9.492$&$162.68$&$ 2.96$&$1.267$&$278.50$&$-0.0050$&$ 0.0155$&$ 1.054$&$ 0.44$&$-0.091$\cr
        1486.02&$   0.049$&$ 0.249$&$  3.70$&$ 0.69$&$0.043$&$  0.49$&$ 0.0035$&$ 0.0051$&$ 0.252$&$  0.05$&$ 0.030$\cr
 \hline
                                                                                                                        
\hline                                                           \enddata
\end{deluxetable}                                                                                                             

\begin{deluxetable}{lrrrrrrrrrrr}                                                                                           
\tabletypesize{\footnotesize}
\tablewidth{470pt}
\tablecaption{\label{tab:modelzero} Light Curve Models: Zero Blending.  }                                                      
\vskip 1em                                                     
\startdata
Model & {$t_0$} & {$u_0$} & {$t_{\rm E}$} & {$q$} & {$Q:\gamma$} & {$\phi$} & {$\pi_{{\rm E},N}$} & {$\pi_{{\rm E},E}$} 
& {$\omega:\alpha_b^{1/2}$} & {$\dot b/b$} \\                                                                           
$\chi^2$    & (day) &$(\times 100)$& (day) &  &$(\times 100)$& (deg) & & & (yr$^{-1}$)& (yr$^{-1}$) \\ \hline           
\hline                                                                                                                  
C-$\parallel $a&$3593.751$&$-8.755$&$173.72$&$ 2.77$&$0.984$&$ 21.69$&$-0.0347$&$ 0.0316$&$ 0.075$&$ 5.59$\cr
  {\bf 1455.38}&$   0.039$&$ 0.028$&$  0.88$&$ 0.32$&$0.018$&$  1.15$&$ 0.0027$&$ 0.0048$&$ 0.047$&$  0.29$\cr
 \hline
C+$\parallel $a&$3593.648$&$ 8.728$&$171.75$&$ 1.42$&$0.940$&$ 14.45$&$-0.0063$&$ 0.0245$&$-0.191$&$ 3.21$\cr
        1474.91&$   0.041$&$ 0.025$&$  0.58$&$ 0.24$&$0.019$&$  1.51$&$ 0.0037$&$ 0.0042$&$ 0.039$&$  0.59$\cr
 \hline
C-$\parallel $b&$3593.612$&$-8.708$&$173.51$&$ 0.85$&$0.898$&$  9.17$&$ 0.0026$&$ 0.0300$&$ 0.368$&$ 1.04$\cr
  {\bf 1463.48}&$   0.028$&$ 0.020$&$  0.48$&$ 0.18$&$0.017$&$  1.68$&$ 0.0021$&$ 0.0046$&$ 0.030$&$  0.59$\cr
 \hline
C+$\parallel $b&$3593.687$&$ 8.711$&$172.11$&$ 0.93$&$0.924$&$ 10.57$&$-0.0008$&$ 0.0180$&$-0.172$&$ 1.39$\cr
        1469.51&$   0.034$&$ 0.021$&$  0.45$&$ 0.11$&$0.016$&$  1.08$&$ 0.0016$&$ 0.0030$&$ 0.028$&$  0.29$\cr
 \hline
C-$\parallel $c&$3593.488$&$-8.769$&$175.52$&$ 0.28$&$0.813$&$ -3.31$&$ 0.0103$&$ 0.0074$&$ 0.753$&$-3.89$\cr
  {\bf 1464.10}&$   0.038$&$ 0.023$&$  0.63$&$ 0.03$&$0.017$&$  1.34$&$ 0.0019$&$ 0.0039$&$ 0.062$&$  0.47$\cr
 \hline
C+$\parallel $c&$3593.493$&$ 8.770$&$174.27$&$ 0.27$&$0.809$&$ -3.67$&$ 0.0084$&$ 0.0040$&$-0.735$&$-3.88$\cr
        1467.54&$   0.038$&$ 0.023$&$  0.67$&$ 0.03$&$0.017$&$  1.22$&$ 0.0015$&$ 0.0030$&$ 0.065$&$  0.43$\cr
 \hline
C+$\parallel $d&$3593.751$&$ 8.725$&$172.80$&$ 0.65$&$0.919$&$  7.20$&$ 0.0025$&$ 0.0115$&$-0.137$&$-0.11$\cr
        1470.98&$   0.051$&$ 0.027$&$  0.50$&$ 0.10$&$0.015$&$  1.47$&$ 0.0015$&$ 0.0041$&$ 0.047$&$  0.55$\cr
 \hline
C-$\perp     $a&$3593.233$&$-8.524$&$166.86$&$ 1.77$&$0.762$&$279.76$&$-0.0168$&$ 0.0310$&$ 0.746$&$ 0.73$\cr
  {\bf 1460.01}&$   0.031$&$ 0.019$&$  0.49$&$ 0.23$&$0.030$&$  0.48$&$ 0.0020$&$ 0.0049$&$ 0.055$&$  0.25$\cr
 \hline
C+$\perp     $a&$3593.269$&$ 8.524$&$166.68$&$ 1.56$&$0.799$&$279.87$&$-0.0119$&$ 0.0211$&$-0.597$&$ 0.73$\cr
        1470.98&$   0.032$&$ 0.019$&$  0.50$&$ 0.25$&$0.034$&$  0.49$&$ 0.0017$&$ 0.0039$&$ 0.058$&$  0.22$\cr
 \hline
W+$\parallel $a&$3593.590$&$ 8.522$&$167.60$&$ 8.33$&$0.953$&$  8.38$&$ 0.0062$&$ 0.0092$&$ 0.267$&$-0.21$\cr
        1479.54&$   0.059$&$ 0.020$&$  0.46$&$ 6.18$&$0.016$&$  0.78$&$ 0.0021$&$ 0.0035$&$ 0.161$&$  0.20$\cr
 \hline
W-$\parallel $a&$3593.718$&$-8.449$&$170.56$&$ 2.40$&$0.937$&$  6.51$&$ 0.0088$&$ 0.0033$&$ 0.296$&$ 0.49$\cr
        1471.40&$   0.033$&$ 0.040$&$  0.71$&$ 0.38$&$0.014$&$  0.44$&$ 0.0042$&$ 0.0032$&$ 0.175$&$  0.02$\cr
 \hline
W+$\parallel $b&$3593.709$&$ 8.427$&$170.01$&$ 2.41$&$0.931$&$  6.54$&$ 0.0055$&$ 0.0021$&$ 0.210$&$ 0.50$\cr
        1471.49&$   0.034$&$ 0.037$&$  0.90$&$ 0.37$&$0.014$&$  0.44$&$ 0.0030$&$ 0.0028$&$ 0.163$&$  0.02$\cr
 \hline
W-$\perp     $a&$3593.552$&$-8.742$&$173.11$&$ 2.20$&$1.169$&$279.61$&$-0.0051$&$ 0.0179$&$ 0.880$&$-0.73$\cr
        1474.83&$   0.067$&$ 0.028$&$  1.11$&$ 0.97$&$0.038$&$  0.59$&$ 0.0030$&$ 0.0055$&$ 0.398$&$  0.25$\cr
 \hline
W+$\perp     $a&$3593.642$&$ 8.759$&$172.88$&$ 4.51$&$1.115$&$280.27$&$-0.0010$&$ 0.0126$&$ 0.769$&$-0.76$\cr
        1480.41&$   0.069$&$ 0.025$&$  1.15$&$ 2.76$&$0.036$&$  0.62$&$ 0.0025$&$ 0.0051$&$ 0.359$&$  0.21$\cr
 \hline
                                                                                                                        
\hline                                                                                                                  
\enddata
\end{deluxetable}    

\clearpage

\clearpage
\thispagestyle{empty}
\clearpage

\end{document}